\documentclass[aps,prd,preprintnumbers,showpacs,nofootinbib,onecolumn]{revtex4}
\usepackage{graphicx}
\usepackage{bm}
\usepackage{amsmath}
\usepackage{amssymb}
\usepackage{amsthm}
\usepackage{color}

\newcommand{\diff}{{\rm d}}
\newcommand{\A}{{\mathcal A}}

\newcommand{\HH}{{\mathcal H}}
\newcommand{\K}{\vec{k}}
\newcommand{\be}{\begin{equation}}

\newcommand{\rr}{\hat{r}}

\newcommand{\ee}{\end{equation}}
\newcommand{\Ft}{\tilde{F}}
\newcommand{\lsim}   {\mathrel{\mathop{\kern 0pt \rlap
  {\raise.2ex\hbox{$<$}}}
  \lower.9ex\hbox{\kern-.190em $\sim$}}}
\newcommand{\gsim}   {\mathrel{\mathop{\kern 0pt \rlap
  {\raise.2ex\hbox{$>$}}}
  \lower.9ex\hbox{\kern-.190em $\sim$}}}

\begin{document}

\title{Stability of Horndeski vector-tensor interactions}

\author{Jose Beltr\'an Jim\'enez$^a$, Ruth Durrer$^b$, Lavinia Heisenberg$^{b,c}$, Mikjel Thorsrud$^d$}

\affiliation{$^a$Centre for Cosmology, Particle Physics and Phenomenology,
Institute of Mathematics and Physics, Louvain University,
2 Chemin du Cyclotron, 1348 Louvain-la-Neuve, Belgium}
\affiliation{$^b$D\'epartement de Physique Th\'eorique and Center for Astroparticle Physics,
Universit\'e de Gen\`eve, 24 quai Ansermet, CH--1211 Gen\`eve 4,
Switzerland}
\affiliation{$^c$Department of Physics, Case Western Reserve University,
10900 Euclid Ave, Cleveland, OH
44106, USA}
\affiliation{$^d$Institute of Theoretical Astrophysics, University of Oslo,
P.O. Box 1029 Blindern, N-0315 Oslo, Norway}
\date{\today}

\begin{abstract}
We study the Horndeski vector-tensor theory that leads to second order equations of motion and contains a non-minimally coupled  abelian gauge vector field. This theory is remarkably simple and consists of only 2 terms for the vector field, namely: the standard Maxwell kinetic term and a coupling to the dual Riemann tensor.  Furthermore, the vector sector respects the $U(1)$ gauge symmetry and the theory contains only one free parameter, $M^2$, that controls the strength of the non-minimal coupling. 
We explore the theory in a de Sitter spacetime and study the presence of instabilities and show that it corresponds to an attractor solution in the presence of the vector field. We also investigate the cosmological evolution and stability of perturbations in a general FLRW spacetime.  We find that a sufficient condition for the absence of ghosts is  $M^2>0$. Moreover, we study further constraints coming from imposing the absence of Laplacian instabilities. Finally, we study the stability of the theory in static and spherically symmetric backgrounds (in particular, Schwarzschild and Reissner-Nordstr\"om-de Sitter). We find that the theory, quite generally, do have ghosts or Laplacian instabilities in regions of spacetime where the non-minimal interaction dominates over the Maxwell term. 
We also calculate the propagation speed in these spacetimes and show that superluminality is a quite generic phenomenon in this theory. 

\end{abstract}

\maketitle

\section{Introduction}
One of the challenging puzzles in modern Cosmology remains the late time accelerated expansion of the Universe. Although the introduction of a simple cosmological constant can account for the cosmic expansion, it is very unsatisfactory since it corresponds to a vacuum energy density of about $(10^{-3}$eV$)^4$ and we have no way of understanding such a small vacuum energy. 

This fact has motivated many attempts to explain late time acceleration by resorting to new dynamical degrees of freedom, being the scalar fields the most widely studied case. In this context, theories like $f(R)$, Horndeski, Galileon and Massive Gravity have been developed and received much attention recently \cite{Sotiriou:2008rp, Horndeski, Nicolis:2008in,deRham:2010ik}. In the context of the most general scalar-tensor theory with non-minimal couplings to curvature and leading to second order equations of motion, both for the field and the metric tensor, the precise allowed tensor structure was first discussed by Horndeski~\cite{Horndeski}. Hereby, having second order equations of motion ensures the absence of extra degrees of freedom which could potentially exhibit the Ostrogradsky instability~\cite{Ostrogradski}. 

Interestingly, the Galilean invariant interactions can be regarded as the effective terms arising from the projection on the brane of Lovelock invariants in a bulk with Poincar\'e invariance in 5-D so that the Galilean symmetry is just the corresponding {\it non-relativistic limit} in which the speed of the brane is small \cite{deRham:2010eu}. Also, in scenarios with extra dimensions, Galileon type of interactions can be obtained from a Kaluza-Klein reduction of Lovelock invariants in higher dimensions \cite{VanAcoleyen:2011mj}. These approaches have the advantage that one automatically obtains the required non-minimal couplings to curvature that yield second order equations of motion and whose minimal terms can be computed directly  to compensate the higher order derivatives of the metric tensor that appear in the energy-momentum tensor \cite{Deffayet:2009wt}.  It is worth mentioning that Galileon-like interactions also arise in a natural manner in  ghost-free massive gravity theories \cite{deRham:2010ik}. Since their inception, there have been many investigations of their cosmological consequences \cite{galileoncosmology}, generalizations to include several species \cite{Multigalileons} or  additional internal gauge symmetries \cite{gaugegalileons}.

Extensions of the Galileon interactions were generalized to the case of arbitrary $p$-forms in an arbitrary number of dimensions in \cite{pGalileons} and a systematic investigation of the phenomenology of models based on this construction including vector fields has been recently performed in \cite{Tasinato:2013oja}. Already Horndeski worked out the most general non-minimal coupling of a Maxwell field yielding second order equations of motion even though higher derivative terms are present in the action \cite{Horndeski:1976gi}. In a similar way as for the Galileons, these general vector-tensor interactions can also be obtained from a Kaluza-Klein reduction of 5-dimensional Gauss-Bonnet terms \cite{KKreduction}.

One motivation to explore higher spin fields in a cosmological context is the discovery of some observed anomalies,  comprising the alignment of the low multipoles of the CMB \cite{Land:2005ad} or the hemispherical asymmetry \cite{asymmetry} (see  also ~\cite{Planck}). These observations might be signaling a potential violation of the cosmological principle by the existence of a privileged direction in the universe that could be related to the presence of a vector field. On the other hand, vector fields  have been used to drive inflation, being the first attempted performed in \cite{Ford}. More recently, an inflationary epoch driven by a set of massive vector fields has been proposed in \cite{triad}. 

Vector fields have also been considered as candidates to explain the current phase of accelerated expansion \cite{vectorDE} and they could even help alleviating the naturalness problem \cite{Cosmicvector} that are present in other dark energy models. Often they can give rise to interesting and non-trivial phenomenologies which have been proven impossible to realize in the context of scalar-tensor theories \cite{Gomes:2013ema}. This includes inflationary solutions with anisotropic hair \cite{f2F2} and cosmologies where a matter era and a scaling attractor with $70\%$ ($30\%$) dark energy (matter) is joined in a single solution \cite{Thorsrud:2012mu}. In \cite{Carlesi:2011gv}, it was shown that the presence of a cosmic vector field could also explain the observation of high redshift massive clusters.  On the other hand, the CMB imposes constraints on observationally viable cosmological models in which only the spatial components are dynamical, because the presence of anisotropic sources could conflict with the observed isotropy. However, it has been shown recently~\cite{Cembranos:2012kk} that, although the energy-momentum tensor of a vector field with a potential is in general anisotropic, its corresponding energy-momentum tensor becomes isotropic when averaging over rapid oscillations as compared to the Hubble expansion rate. Therefore, the spatial components may actually be harmless concerning the generation of large anisotropy in the cosmological evolution.  Also a screening mechanism for conformally coupled vector fields has been studied in \cite{Jimenez:2012ph}.

There are few known examples of fully viable vector theories besides the standard gauge invariant Maxwell action. Notable exceptions to this include couplings of the Maxwell term to a positive definite function of a scalar field, which have extensively been studied in the context of inflation recently in \cite{f2F2}  (see also references therein). Also a parity violating coupling of a function of the inflaton to the electromagnetic field has been explored as a potential mechanism for the generation of helical magnetic fields \cite{Durrer:2010mq}. Quite generically, other Lagrangians built from contractions of vector fields (with itself or with other fields) are plagued by instabilities, both classically and quantum mechanically \cite{vectorInstability}, although some room might remain \cite{viability}. These problems stem from additional degrees of freedom arising from the lost gauge symmetry or the resulting higher order field equations \cite{EspositoFarese:2009aj}.  Horndeski's vector-tensor interaction, on the other hand, respects the gauge symmetry and yields second order field equations.  Although promising, this is not enough to guarantee a stable theory; therefore, the main purpose of this paper is to investigate the stability conditions of this theory by studying the presence of ghosts and/or Laplacian instabilities in several physically relevant geometries. 

The paper is organized in the following way. In section \ref{ch:theory} we present and discuss the theory and the corresponding equations. In the following section we review the cosmological dynamics of the vector. Then, in section \ref{hamiltonian}, we investigate the Hamiltonian stability by identifying conditions for avoidance of ghosts and Laplacian instabilities in a number of classical spacetime backgrounds.  We also calculate the propagation speed of the vector field. In section \ref{discussion} we conclude by summarizing our main findings.

In this paper we use $M_p^{-2}=8\pi G$ and the following signature conventions for the metric, Ricci and Riemann tensors: $\text{sign}(g_{\mu\nu})=(+,-,-,-)$, $R_{\mu\nu}=R^\sigma_{\;\;\mu\sigma\nu}$, $R^{\alpha}_{\;\;\beta\gamma\delta}=\partial_\gamma \Gamma^\alpha_{\;\;\beta\delta}-\partial_\delta \Gamma^\alpha_{\;\;\beta\gamma} + \Gamma^\alpha_{\;\;\mu\gamma} \Gamma^\mu_{\;\;\beta\delta} - \Gamma^\alpha_{\;\;\mu\delta} \Gamma^\mu_{\;\;\beta\gamma}$. 

\section{The theory \label{ch:theory}}

Our aim is to study a general action for a vector field with a non-minimal coupling to gravity leading to second order equations of motion for both, the vector field and the gravitational sector. We are interested in an action containing only kinetic terms for the vector field so that neither potential terms nor direct couplings of the vector field to curvature will be considered, but only derivative couplings. 
In principle, we could consider all possible terms involving a coupling of $F_{\mu\nu}$ to the Riemann tensor. However, in order to guarantee that the gravitational equations remain of second order, the couplings must be to a divergence-free tensor constructed with the Riemann tensor. In 4-dimensions, in addition to  the metric there are only two tensors satisfying this condition, namely the Einstein tensor and the dual Riemann tensor
\begin{eqnarray}
G_{\mu\nu}&=&R_{\mu\nu}-\frac{1}{2}Rg_{\mu\nu},\\
L^{\mu\alpha\nu\beta}&=&2R^{\mu\alpha\nu\beta}+2(R^{\mu\beta}g^{\nu\alpha}+R^{\nu\alpha}g^{\mu\beta}-R^{\mu\nu}g^{\alpha\beta}-R^{\alpha\beta}g^{\mu\nu})+R(g^{\mu\nu}g^{\alpha\beta}-g^{\mu\beta}g^{\nu\alpha}).
\end{eqnarray}
The divergence less tensor ${L^\mu}_{\alpha\nu\beta}$ is the dual of the tensor valued curvature form 
${\Omega^\mu}_\alpha = \frac{1}{2}{R^\mu}_{\alpha\nu\beta}\diff x^\nu\wedge \diff x^\beta$  defined by
$L^{\alpha\beta\gamma\delta}= -\frac{1}{2}\epsilon^{\alpha\beta\mu\nu}\epsilon^{\gamma\delta\rho\sigma}R_{\mu\nu\rho\sigma}$, where $\epsilon_{\mu \nu \alpha\beta}=\epsilon_{[\mu \nu \alpha\beta]}$ is the Levi-Civita tensor. This divergenceless tensors are related to the non-trivial Lovelock invariants in 4 dimensions, i.e., the Ricci scalar and the Gauss-Bonnet term.

Since the Einstein tensor is symmetric, its contraction with $F_{\mu\nu}$ vanishes for symmetry reasons $G_{\mu\nu}F^{\mu\nu}=0$. Note that even though $G^{\mu\nu}$ and $g^{\alpha\beta}$ are both divergence-free, their product $G^{\mu\nu}g^{\alpha\beta}$ is not so that we cannot allow the term $G^{\mu\nu}g^{\alpha\beta}F_{\mu\alpha}F_{\nu\beta}$. The same argument applies to the product of two Einstein tensors so that $G^{\mu\nu}G^{\alpha\beta}F_{\mu\alpha}F_{\nu\beta}$ is not allowed either. Only the dual Riemann tensor is divergence-free and it can be coupled to $F_{\mu\alpha}F_{\nu\beta}$.  Thus, the desired action with only second order equations of motion is remarkably simple and reads
\begin{align}
S&=\int \diff^4x\sqrt{-g}\left[-\frac{1}{2}M_p^2R-\frac14F_{\mu\nu}F^{\mu\nu}+\frac{1}{4M^2}L^{\alpha\beta\gamma\delta}F_{\alpha\beta}F_{\gamma\delta} \right]\nonumber \\
&=\int \diff^4x\sqrt{-g}\left[-\frac{1}{2}M_p^2R-\frac14F_{\mu\nu}F^{\mu\nu}+\frac{1}{2M^2}\Big(R F_{\mu\nu}F^{\mu\nu} - 4R_{\mu\nu}F^{\mu\sigma}F^{\nu}_{\;\;\sigma} + R_{\mu\nu\alpha\beta} F^{\mu\nu} F^{\alpha\beta} \Big) \right],
\label{action}
\end{align}
where $M^2$ is the only free parameter of the theory and its sign will be fixed by stability requirements\footnote{In this action one could in principle add an additional parameter in front of the Maxwell parameter. However, in order to recover a stable theory in flat spacetime, such a parameter needs to be negative and, by canonically normalizing the vector field, we can fix it to the usual $-1/4$. }. One might further wonder if interactions of the form $L^{\alpha\beta\gamma\delta}\tilde F_{\alpha\beta}\tilde F_{\gamma\delta}$ and $L^{\alpha\beta\gamma\delta} F_{\alpha\beta}\tilde F_{\gamma\delta}$ (with  $\tilde{F}^{\alpha\beta}=\frac{1}{2}\epsilon^{\alpha\beta\mu\nu}F_{\mu\nu}$ being the dual of $F_{\mu\nu}$) could fulfill our requirements and be a valid interaction which should be taken into account. The latter one would explicitly break the parity invariance, but this symmetry is not one of our requirements. Nevertheless, at a closer look one realizes that these interactions are nothing else but coupling of the Riemann tensor to $F_{\mu\alpha}F_{\nu\beta}$. Since the Riemann tensor is not divergence-free, these interactions give rise to higher order equations of motion. So interactions of the form $L^{\alpha\beta\gamma\delta}F_{\alpha\beta}F_{\gamma\delta}$ fulfill our requirement of second order equation of motion and gauge invariance and are equivalent to $R^{\alpha\beta\gamma\delta}\tilde F_{\alpha\beta}\tilde F_{\gamma\delta}$, while interactions of the form $L^{\alpha\beta\gamma\delta}\tilde F_{\alpha\beta}\tilde F_{\gamma\delta}$ are not since they are equivalent to $R^{\alpha\beta\gamma\delta}F_{\alpha\beta}F_{\gamma\delta}$ and thus give higher order equations of motion.  This reasoning can be understood from the fact that $F_{\mu\nu}$ is a closed form (so it satisfies Bianchi identities) whereas its dual is not (as we shall discuss below). Furthermore, one might wonder whether the interaction changes if we contract the indices between the dual Riemann tensor and $F_{\mu\alpha}F_{\nu\beta}$ in a different way, i.e. whether $L^{\alpha\gamma\beta\delta}F_{\alpha\beta} F_{\gamma\delta}$ gives rise to a different interaction but it turns out to be that this term is proportional to $L^{\alpha\beta\gamma\delta}F_{\alpha\beta} F_{\gamma\delta}$ such that one can reabsorb its effect into $M^2$. Note also that  this Horndeski interaction is the only interaction yielding second order equations of motion in four dimensions. In higher dimensions one can construct other non-minimal interactions based on the divergence-free tensors in that dimension; more precisely, one will have additional divergence-free tensors associated to the corresponding Lovelock invariants which can then be contracted with the field strength tensor.
The second form of the action, Eq.~(\ref{action}), will be convenient for comparing our results to other theories with non-minimal couplings.  

The tensor structure appearing in the above action has also been discussed in \cite{deRham:2011by} for the helicity-0 interactions as an outcome of massive gravity. After working out the scalar field case, Horndeski \cite{Horndeski:1976gi} proved that this coupling is actually the only non-minimal coupling for the electromagnetic field leading to second order equations of motion and recovering Maxwell theory in flat spacetime. As can be seen from the action, Maxwell theory corresponds to the limit where the spacetime curvature is much smaller than $M^2$. In this limit, the non-minimal coupling is strongly suppressed with respect to the usual Maxwell term. Furthermore, it can be obtained from a Kaluza-Klein reduction of the Gauss-Bonnet terms in 5 dimensions \cite{KKreduction}. However, in such a setup a quartic interaction term $\frac{1}{M^2}[(F_{\alpha\beta}F^{\alpha\beta})^2-2F^\alpha_\beta F^\beta_\gamma F^\gamma_\delta F^\delta_\alpha]$ arises together with the Horndeski interaction. Thus, our results will only apply for this scenario provided the vector field does not affect the space-time geometry, i.e., $F^2\ll R$. This will indeed be the case for most of our results, with the only exception of the dynamical system analysis performed for the de Sitter spacetime stability in section \ref{Sec:dSrstability}. More recently, this non-minimal coupling has also been considered in \cite{EspositoFarese:2009aj} and the cosmology of this kind of interactions has been explored in \cite{Barrow:2012ay}.\footnote{$M^2$ is related to the non-minimal coupling parameter, $\lambda$, used in \cite{Barrow:2012ay} in the following way: $\lambda/M_p^2 = 2/M^2$.} 

It is interesting to note that the non-minimal coupling present in (\ref{action}) involves the same type of terms that are obtained when introducing vacuum polarization corrections from a curved background in standard electromagnetism \cite{Drummond} \footnote{These type of terms were also considered in \cite{widrow} as a possible mechanism for the generation of magnetic fields during inflation. This becomes a possibility because the non-minimal interactions break the conformal invariance of the Maxwell term in 4 dimensions.}. In that case,  the non-minimal couplings are suppressed by the mass of the charged fermion running inside the loops (typically the electron mass in standard QED). Such radiative corrections will also arise in our case if the vector field is coupled to some other charged field of mass $m$. This means that the lightest possible particle charged under our $U(1)$ field should satisfy $m\gg M$ for those quantum corrections not to spoil the Horndeski interaction. On the other hand, one might also wonder whether loop corrections involving gravitons could spoil the Horndeski structure as well, since they will produce radiative corrections of this type (i.e., terms linear in the curvature and quadratic in $F_{\mu\nu}$), but without the appropriate coefficients. However, these corrections will be suppressed by the Planck mass so that, for the action (\ref{action}) to make sense as an effective field theory, we would need to require $M\ll M_p$. This is a safe bound since we know that GR breaks down at the Planck scale anyways.

That the particular combination appearing in (\ref{action}) leads to second order equations of motion can be easily understood. Only the non-minimal coupling contains more than two derivatives so this is the only term that could lead to higher order terms in the equations of motion. In order to show that the equations remain of second order is convenient to write the dual Riemann tensor as $L^{\alpha\beta\gamma\delta}=-\frac{1}{2}\epsilon^{\alpha\beta\mu\nu}\epsilon^{\gamma\delta\rho\sigma}R_{\mu\nu\rho\sigma}$. Then, it becomes apparent why the $M^2$-term will lead to second order equations of motion by virtue of the Bianchi identities for the Riemann tensor and $F_{\mu\nu}$.  For instance, if we perform the variation of the non-minimal interaction with respect to $A_\mu$, the only possible danger terms with higher order derivatives will come from derivatives applying on the Riemann tensor once we do integration by parts $\frac{1}{2}\epsilon^{\alpha\beta\mu\nu}\epsilon^{\gamma\delta\rho\sigma}\nabla_\gamma R_{\mu\nu\rho\sigma}F_{\alpha\beta}\delta A_\delta$. Using the Bianchi identity for the Riemann tensor $R_{\mu\nu[\rho\sigma;\gamma]}=0$ we see that this dangerous terms automatically cancel\footnote{The Bianchi identity for an arbitrary p-form $d\omega=0$, where $\omega$ is the strength of the p-form, guaranties that the Lagrangian $\epsilon^{\mu\nu\dots}\epsilon^{ab\dots}\omega_{\mu\nu\dots}\omega_{ab\dots}\cdots(\partial_\rho\omega_{cd\dots})\cdots(\partial_e\omega_{\sigma\tau\dots})$ for the p-form will only give rise to second-order equations of motion  \cite{pGalileons}.}.  We can also see this  explicitly by computing the corresponding equations of motion. The non-gravitational field equations are therefore  of second order and are given by
\begin{eqnarray}
\left[g^{\mu\rho}g^{\nu\sigma}-\frac{1}{M^2}L^{\mu\nu\rho\sigma}\right]\nabla_\nu F_{\rho\sigma}&=&0 \,.
\end{eqnarray}
Because of the transversality of the dual Riemann tensor, we see that the above equation is divergence-free. Moreover, since  $L^{\alpha\beta\gamma\delta}$ is divergence-free, we can also  write the above equation as
\begin{eqnarray}
\nabla_\nu\left[F^{\mu\nu}-\frac{1}{M^2}L^{\mu\nu\rho\sigma}F_{\rho\sigma}\right]=0,
\end{eqnarray}
which resembles the usual form of Maxwell equations.

Varying the action~(\ref{action}) with respect to the metric yields the following energy momentum tensor for the vector field
\begin{eqnarray}
T_{\mu\nu}&=&-F_{\mu\alpha}F_\nu^{\;\;\alpha}
+\frac{1}{4}g_{\mu\nu}F_{\alpha\beta}F^{\alpha\beta}\nonumber\\
&&+\frac{1}{2M^2}\left[-R_{\alpha\beta\gamma\delta}\tilde{F}^{\alpha\beta}\tilde{F}^{\gamma\delta}g^{\mu\nu}+2 R^\mu_{\;\beta\gamma\delta}\tilde{F}^{\nu\beta}\tilde{F}^{\gamma\delta}+4\nabla_\gamma\nabla_\beta\left(\tilde{F}^{\mu\beta}\tilde{F}^{\gamma\nu}\right)\right].
\label{Tmunu}
\end{eqnarray}
Although the $M^2$-term of the energy momentum tensor might seem to contain more than second order derivatives, because $\tilde{F}^{\alpha\beta}$ is divergence-free in the absence of external currents, this is actually not the case. In fact, it can be written in the more suggestive form
\begin{eqnarray}\label{e:ddff}
\nabla_\gamma\nabla_\beta\left(\Ft^{\mu\beta}\Ft^{\gamma\nu}\right)=R^\mu_{\lambda\beta\gamma}\Ft^{\lambda\beta}\Ft^{\gamma\nu}+R_{\lambda\gamma}\Ft^{\mu\lambda}\Ft^{\gamma\nu}+\nabla_\gamma\Ft^{\mu\beta}\nabla_\beta\Ft^{\gamma\nu} \,.
\end{eqnarray}
If a current is present so that  $\tilde{F}^{\alpha\beta}$ is no longer divergence-free, this equation acquires a contribution from the current.
In Eq.~(\ref{e:ddff})  one sees explicitly that only second derivatives are present.

It is interesting to note that  the energy-momentum tensor given in (\ref{Tmunu}) does not reduce to the conventional expression even in flat Minkowski spacetime where the $M^2$-term gives rise to a contribution proportional to $\partial_\gamma\partial_\beta(\Ft^{\mu\beta}\Ft^{\gamma\nu})$ even though such a term is not present in the action. At first sight this might seem to lead to an inconsistency. To show that this is not the case, we have to notice that this term is of the form $\partial_\beta \Theta^{\beta\mu\nu}$ with $\Theta^{\beta\mu\nu}=\partial_\gamma(\Ft^{\mu\beta}\Ft^{\gamma\nu})$. Furthermore, 
 we recall that, in flat spacetime, the energy-momentum tensor is only defined up to the addition of the divergence of an arbitrary tensor antisymmetric in the first two indices. We can use this freedom to get rid of the term proportional to $M^{-2}$ in the energy-momentum tensor in flat spacetime.  Equivalently, since this term is just a divergence, its Lorentz generators vanish and it does not contribute to physical quantities like the total energy, momentum or angular momentum of the field.
 In curved spacetime, when the Riemann tensor is non-vanishing, this is in general no longer the case.

\section{Cosmological evolution \label{evolution}}
In this section we study the cosmological evolution of a homogenous vector field which has also been considered in \cite{EspositoFarese:2009aj} and \cite{Barrow:2012ay}.  We start by considering the vector field as subdominant (i.e. we neglect its backreaction to the geometry) and study its dynamics on a Friedmann-Lema\^{\i}tre-Robertson-Walker (FLRW) spacetime with special emphasis in matter and radiation dominated epochs. Then, we pursue a dynamical system approach to study the stability of the de Sitter background in the presence of the vector field.  

\subsection{Subdominant homogeneous vector field}\label{sec:test}
We start by considering the evolution of the vector field in a  homogeneous and isotropic  FLRW universe, which can be described by the following metric: 
\begin{equation}
ds^2=a(\eta)^2(d\eta^2-d\vec{x}^2),
\end{equation}
with $\eta$ being the conformal time. We consider only the homogeneous mode for the vector field, although we shall keep all its components. This is incompatible with the symmetries of the FLRW metric, but we assume that the vector field is sufficiently subdominant so that we can neglect its contribution to Einstein equations, i.e. it is a 'test field'. Thus, we have\footnote{The temporal component $A_0$ plays no role in the homogeneous evolution and, in fact, can always be gauged away. On the other hand, the used form for the vector field implies vanishing homogeneous magnetic part. However, since the Bianchi identity for $F_{\mu\nu}$ implies that its magnetic component decays, we do not expect our results to change even if we include such a component.} $\A=(0,\vec{A}(t))$. In this case, the equation of motion for $\A$ can be written as
\begin{eqnarray}
\left(1+\frac{4 \HH^2}{M^2a^2}\right)\vec{A}''-\frac{8\HH\left(\HH^2-\HH'\right)}{M^2a^2}\vec{A}'=0 \,.
\end{eqnarray}
Here $\HH =a'/a =aH$ is the comoving Hubble parameter and $H=a'/a^2$ is the physical Hubble parameter. 
Notice that, for pure de Sitter expansion we have $\HH'=\HH^2$ and, therefore, the dependence on $M^2$ factors out from the equation of $\vec{A}$ and we have simply $\vec{A}''=0$, like for conventional electrodynamics which is conformally invariant. Thus, the vector field in a de Sitter phase evolves like $\vec{A}(t)=\vec{A}_g+\vec{A}_p\eta$ irrespectively of the non-minimal interaction, which does not have any effect\footnote{However, notice that during inflation the expansion is not exactly de Sitter, but a quasi de Sitter expansion characterized by a small slow roll parameter. Hence, in that case there will be a mild dependence on $M^2$.}. We will see this in more detail below. On the other hand,  if we assume a power law expansion with $a\propto \eta^\alpha$, i.e., $\HH=\alpha/\eta$, $\alpha\neq -1$, the solution for the vector field is
\begin{eqnarray}
\vec{A}(t)&=&\vec{A}_g+\vec{A}_p\int\frac{\eta^{2(1+\alpha)}}{4 \alpha^2+M^2\eta^{2+2\alpha}}\diff\eta \,,
\end{eqnarray}
where $\vec{A}_g$ is a pure gauge mode that will play no role and $\vec{A}_p$ some constant amplitude. Notice that this solution can be alternatively expressed as (dropping the gauge mode)
\begin{eqnarray}
\vec{A}(t)&=&\vec{A}_p\int\frac{d\eta}{1+\frac{4\HH^2}{a^2M^2}} \,.
\end{eqnarray}
In a FLRW universe, the corresponding homogeneous energy density obtained from (\ref{Tmunu}) reads
\begin{eqnarray}
\rho_{A}=\frac{1}{2a^4}\left(1+\frac{12 \HH^2}{M^2a^2}\right)\vert\vec{A}'\vert^2=\frac{1}{2a^4}\frac{1+\frac{12 \HH^2}{M^2a^2}}{\left(1+\frac{4\HH^2}{a^2M^2}\right)^2}\vert\vec{A}_p\vert^2 \,,
\label{rhoRW}
\end{eqnarray}
where we have used the solution obtained above. We find two regimes: if $ \frac{\HH^2}{a^2M^2}\gg1$, then $\rho_A \propto (a\HH)^{-2}\propto a^{2/\alpha-2}$ and if $ \frac{\HH^2}{a^2M^2}\ll1$, then $\rho_A\propto 1/a^4$, as usual for a Maxwell field. Thus, in the early universe where the relativistic degrees of freedom dominate the energy content (i.e., $\alpha=1$), $\rho_A$ will be constant until $\HH^2\simeq a^2M^2$ (or equivalently $H^2\simeq M^2$, where $H$ is the Hubble expansion rate in cosmic time) and from that time on, it will decay as a radiation-like fluid. If $M$ is smaller than the Hubble expansion rate at the time of matter-radiation equality, then $\rho_A$ will be initially constant, after equality time ($\alpha=2$) it will decay as $a^{-1}$ and, again when $\HH^2=a^2M^2$ it will start decaying like $a^{-4}$. Finally, notice that during a de Sitter phase the energy density always decays as $a^{-4}$ irrespectively of the ratio $\frac{\HH^2}{a^2M^2}$.

\begin{figure}[h!]
\begin{center} 
\includegraphics[width=11.5cm]{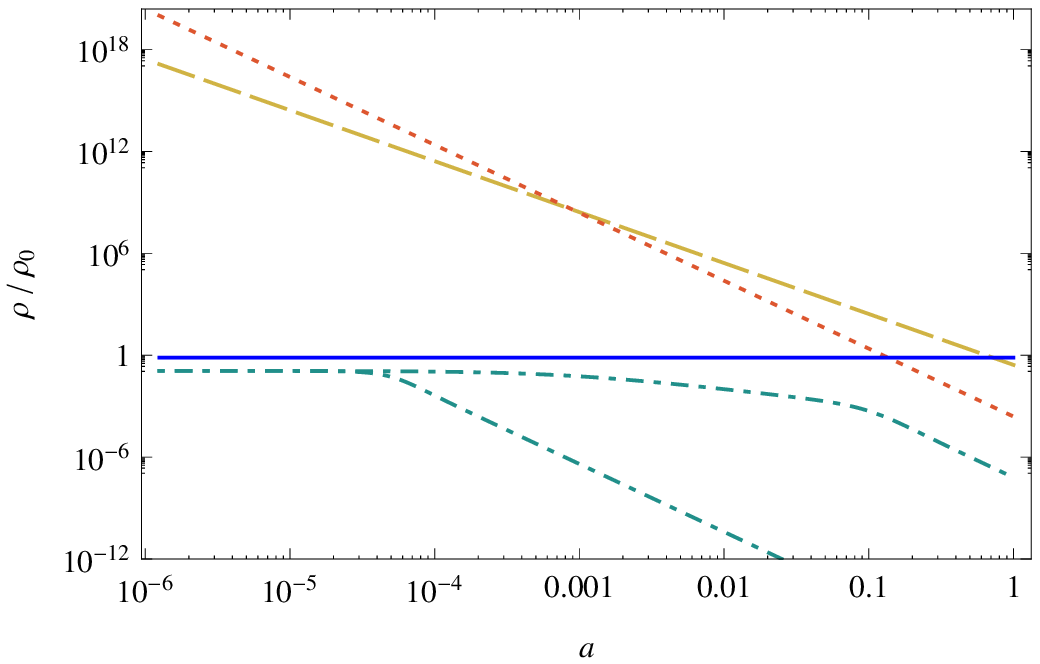}
\includegraphics[width=6cm]{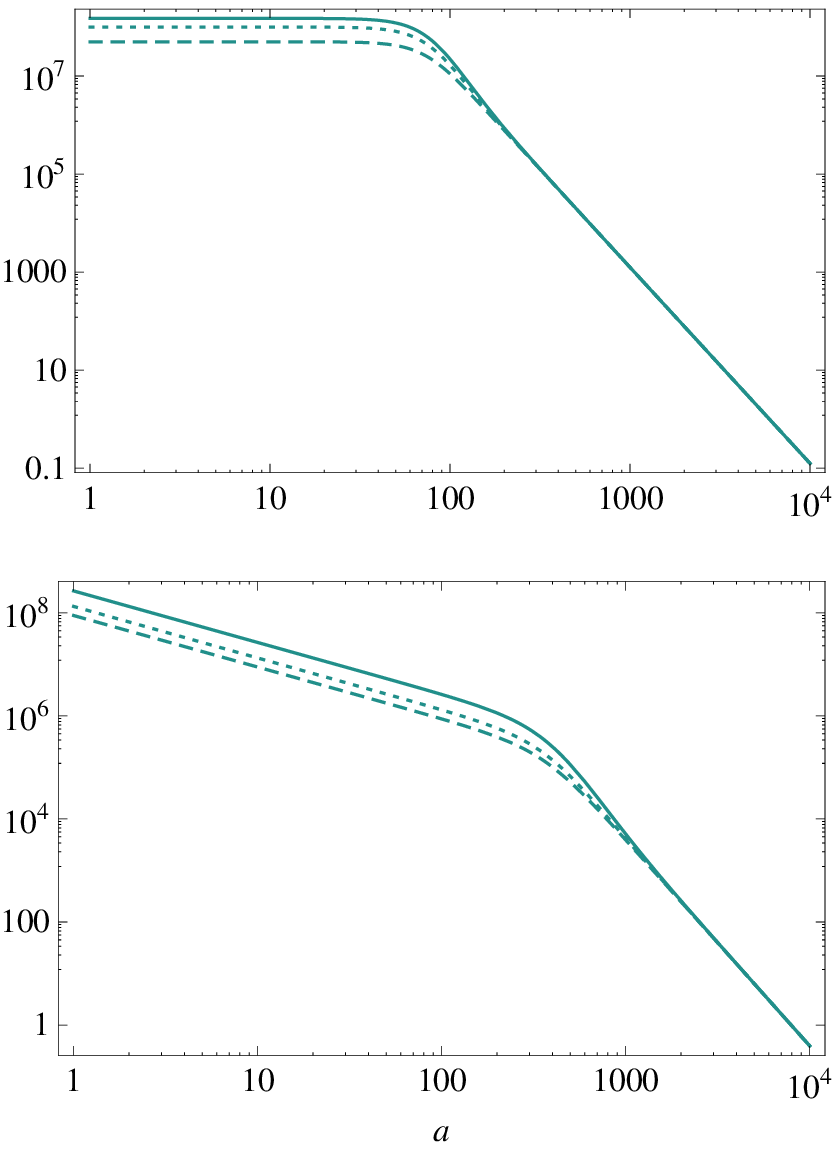}
\caption{Left panel shows the cosmological evolution for the standard components of the universe, namely radiation (dotted) and matter (dashed), as well as the constant energy density of the cosmological constant. We also show the evolution of $\rho_A$ (dotdashed) for $M=10^7H_0$ and $M=25H_0$ as two examples in which the regime with $M^2\gsim H^2$ (i.e. $\rho_A\propto a^{-4}$) is reached in the radiation and matter dominated epochs respectively. On the right panel we show the evolution for the energy density $\rho_A$ (solid) and the transverse (dotted) and longitudinal (dashed) pressures coming from $\vec{A}$ in a radiation dominated  universe (upper panel) and in a matter dominated  universe (lower panel). The pressures evolve in the same way as $\rho_A$. }
\label{cosmoevol}
\end{center}
\end{figure}

We now want to study the generation of large scale anisotropy in this model. For this we need to compute the anisotropic stress. If we choose $\vec{A}$ pointing along the $z$-direction, the longitudinal and transverse pressures are given by:
\begin{align}
&p_\parallel=-T^z_z=-\frac{1}{2a^4}\left(1+\frac{4 \HH^2}{M^2a^2}\right)(A'_z)^2,\\
&p_\perp=-T^x_x=+\frac{1}{2a^4}\left[\left(1-4\frac{ \HH'-3\HH^2}{M^2a^2}\right)(A'_z)^2-\frac{8 \HH}{M^2a^2}A'_zA''_z\right].
\end{align}
Again we have two different regimes depending on the ratio $\frac{\HH^2}{a^2M^2}$ as with $\rho_A$. Moreover, the corresponding behaviour for the pressures is also the same as for $\rho_A$ (see Fig. \ref{cosmoevol}). Thus, the only possibility to have a non-negligible contribution to the large scale anisotropy of the universe is $M\lsim H_{eq}$ with $H_{eq}$ the Hubble expansion rate at matter and radiation equality. If the value of $M$ is larger than $H_{eq}$, the contribution coming from $\vec{A}$ will always be negligible because if it was initially subdominant with respect to radiation, it will always remain subdominant (see left panel of Fig. \ref{cosmoevol}).  In any case, a safe constraint on $M$ in order not to spoil the CMB quadrupole measurement is $M\gsim H_{eq}\simeq10^{-29}$ eV, which is a very weak constraint.

\subsection{Stability of the de Sitter universe}\label{Sec:dSrstability}

In the previous section we have studied the cosmological evolution when the universe is dominated by a component with constant equation of state and the vector field contributes negligibly to the expansion. Now we shall study the case in which the universe is dominated by the vector field plus a cosmological constant and we shall show that the de Sitter universe is an attractor. Because of the presence of the vector field, the isotropic solution is not exact so that we need to study the stability of the de Sitter solution under small perturbations of the vector field. For that, we assume homogeneous fields again, with the vector field pointing along the $z$-direction, in an axisymmetric Bianchi I metric described by the following line element:
\be
ds^2=dt^2-a^2_\perp(t)\left(dx^2+dy^2\right)-a^2_\parallel(t) dz^2,
\ee
which is compatible with the existing symmetries. On this spacetime, the vector field equations read
\begin{eqnarray}\label{eom:A}
\left(1+\frac{4H_\perp^2}{M^2}\right)\ddot{\vec{A}}+\left[\left(1+\frac{4 H_\perp^2}{M^2}\right)(2H_\perp-H_\parallel)+\frac{8 H_\perp\dot{H}_\perp}{M^2}\right]\dot{\vec{A}}&=&0,
\end{eqnarray}
where $H_{\perp}=\dot{a}_\perp/a_\perp$ and $H_{\parallel}=\dot{a}_\parallel/a_\parallel$ are the expansion rates along the transverse and longitudinal directions respectively. For the isotropic case with $H_\perp=H_\parallel=H$ we recover the equations of the previous section. In this case the Einstein equations can be written as
\begin{eqnarray} \label{e:const}
H_\perp^2+2H_\parallel H_\perp&=&8\pi G\rho,\\ \label{e:Eperp}
\dot{H}_\perp+\dot{H}_\parallel+H_\perp^2+H_\parallel^2+H_\perp H_\parallel&=&-8\pi Gp_\perp,\\
2\dot{H}_\perp+3H_\perp^2&=&-8\pi Gp_\parallel,    \label{e:Epap}
\end{eqnarray}
with
\begin{eqnarray}
\rho&=&\rho_\Lambda+\frac{1}{2a_\parallel^2}\left(1+\frac{12 H_\perp^2}{M^2}\right)\dot{A}_z^2,\\
p_\perp&=&-\rho_\Lambda+\frac{1}{2a_\parallel^2}\left[\left(
1-4\frac{H_\perp(H_\perp-H_\parallel)+\dot{H}_\perp}{M^2}
\right)\dot{A}_z^2-\frac{8 H_\perp}{M^2}\dot{A}_z\ddot{A}_z\right],\\
p_\parallel&=&-\rho_\Lambda-\frac{1}{2a_\parallel^2}\left(1+\frac{4 H_\perp^2}{M^2}\right)\dot{A}_z^2.
\end{eqnarray}
To proceed to the analysis of the equations and study the corresponding critical points, it will be convenient to split the two expansion rates as follows:
\begin{eqnarray}
H_\perp=H(1-R),\\
H_\parallel=H(1+2R).
\end{eqnarray}
Here, $H$ represents the average expansion rate defined as follows: 
\be
a=(a_\perp^2a_\parallel)^{1/3}\,, \quad H =\frac{\dot a}{a} = \frac{2}{3} H_\perp +  \frac{1}{3} H_\parallel \,.
\ee
 The Hubble normalized shear, $R$, measures the deviation from the isotropic case. Since $\rho_\Lambda$ is constant, we can eliminate $a_\parallel^{-2}\dot{A}_z^2$ using the constraint equation (\ref{e:const}) and $\ddot A_z$ using the equation of motion  (\ref{eom:A}). We can then combine Eqs.~ (\ref{e:Eperp}) and (\ref{e:Epap})  to an autonomous system of first order differential equations for $H$ and $R$. A somewhat lengthy algebra yields
\begin{eqnarray} 
\dot H&=&\frac{1}{3g}\left(\frac{2\rho_\Lambda}{M_p^2}+f^H_2H^2+f^H_4\frac{H^4}{M^2}+f^H_6\frac{H^6}{M^4}+f^H_8\frac{H^8}{M^6}\right),\\
\dot R&=&\frac{(-1+R)}{9Hg}\left(-\frac{6\rho_\Lambda}{M_p^2}+f^R_2H^2+f^R_4\frac{H^4}{M^2}+f^R_6\frac{H^6}{M^4}+f^R_8\frac{H^8}{M^6}\right),
\end{eqnarray} 
where
\begin{eqnarray}
f^H_2&=&-3(2+R^2)+8(1-R)\left[(7-10R)\frac{\rho_\Lambda}{M_p^2M^2}-2(1-R)\frac{\rho_\Lambda^2}{M_p^4M^4}\right],\\
f^H_4&=&12(1-R)^2\left[-14+R(6-R)+(1-R)^2(\frac{48\rho_\Lambda}{M_p^2M^2}+\frac{16\rho_\Lambda^2}{M_p^4M^4})\right],\\
f^H_6&=&144(1-R)^4\left[-11+2R(1-R)(1-\frac{4\rho_\Lambda}{M_p^2M^2})\right],\\
f^H_8&=&1728(1-R)^6(1+2R^2),
\end{eqnarray}
and similarly
\begin{eqnarray}
f^R_2&=&9(1+R)(2+R)-12(1-R)(11-20R)\frac{\rho_\Lambda}{M_p^2M^2}+48(1-R)^2\frac{\rho_\Lambda^2}{M_p^4M^4},\\
f^R_4&=&36(1-R)^2\left[11+R(12+R)-16(1-R)(2-3R)\frac{\rho_\Lambda}{M_p^2M^2}-16(1-R)^2\frac{\rho_\Lambda^2}{M_p^4M^4}\right],\\
f^R_6&=&432(1-R)^4\left[7+R(9+2R)+4(1-R)(1+2R))\frac{\rho_\Lambda}{M_p^2M^2}\right],\\
f^R_8&=&10368R(1-R)^6(1+R),
\end{eqnarray}
with $g=\left[1+4(1-R)^2\frac{H^2}{M^2}\right]\left[1+12(1-R)^2\frac{H^2}{M^2}\right]^2$ and $M^{-2}_p=8\pi G$. 
This autonomous system has two critical points given by
\begin{eqnarray}
&{\rm I}:&\;\;\; \left(\;R=0,\;\; H=\sqrt{\frac {\rho_\Lambda}{3M_p^2}}\right),\\
&{\rm II}:&\;\;\;\left(\;R=1,\;\; H=\frac{\sqrt{2\rho_\Lambda}}{3M_p}\right).
\end{eqnarray}

\begin{figure}
\begin{center} 
\includegraphics[width=5.75cm]{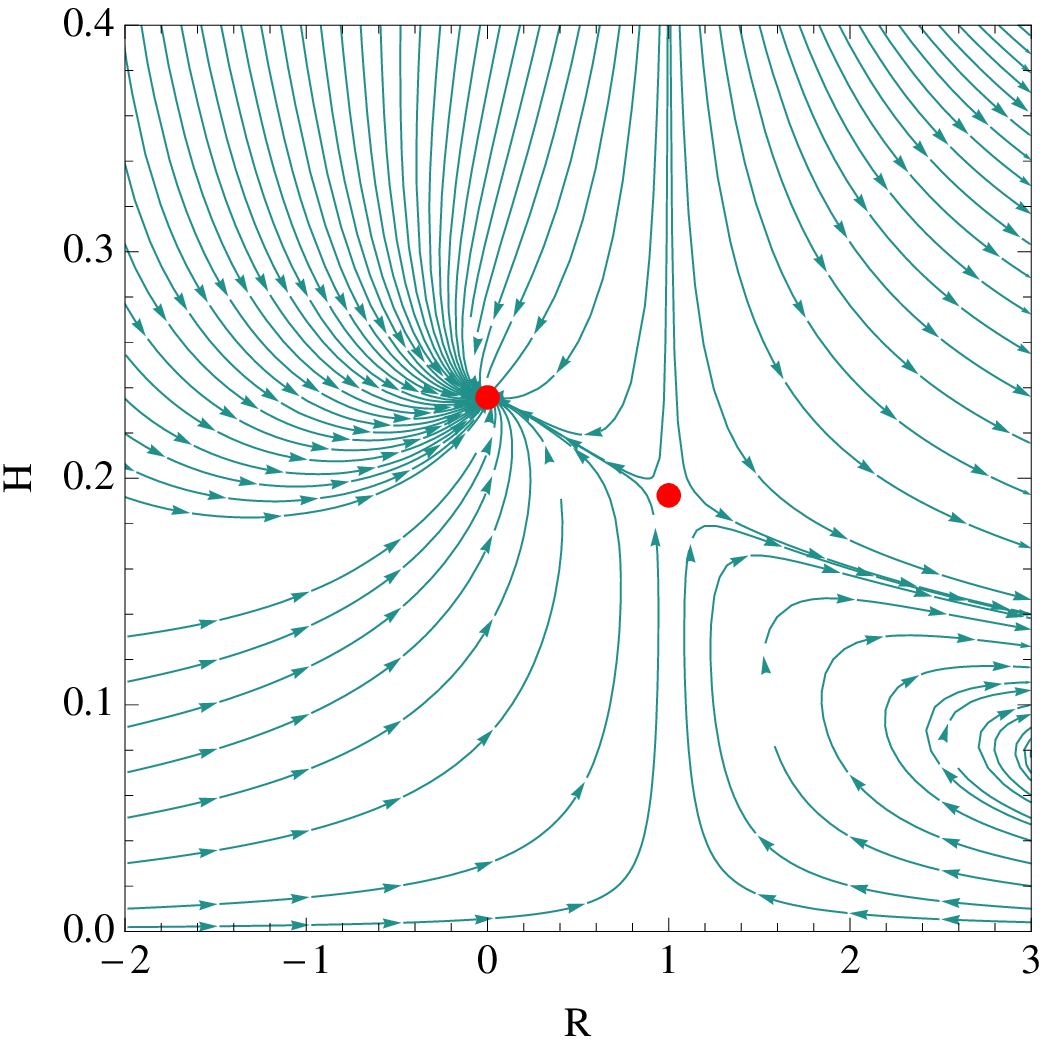}
\includegraphics[width=5.75cm]{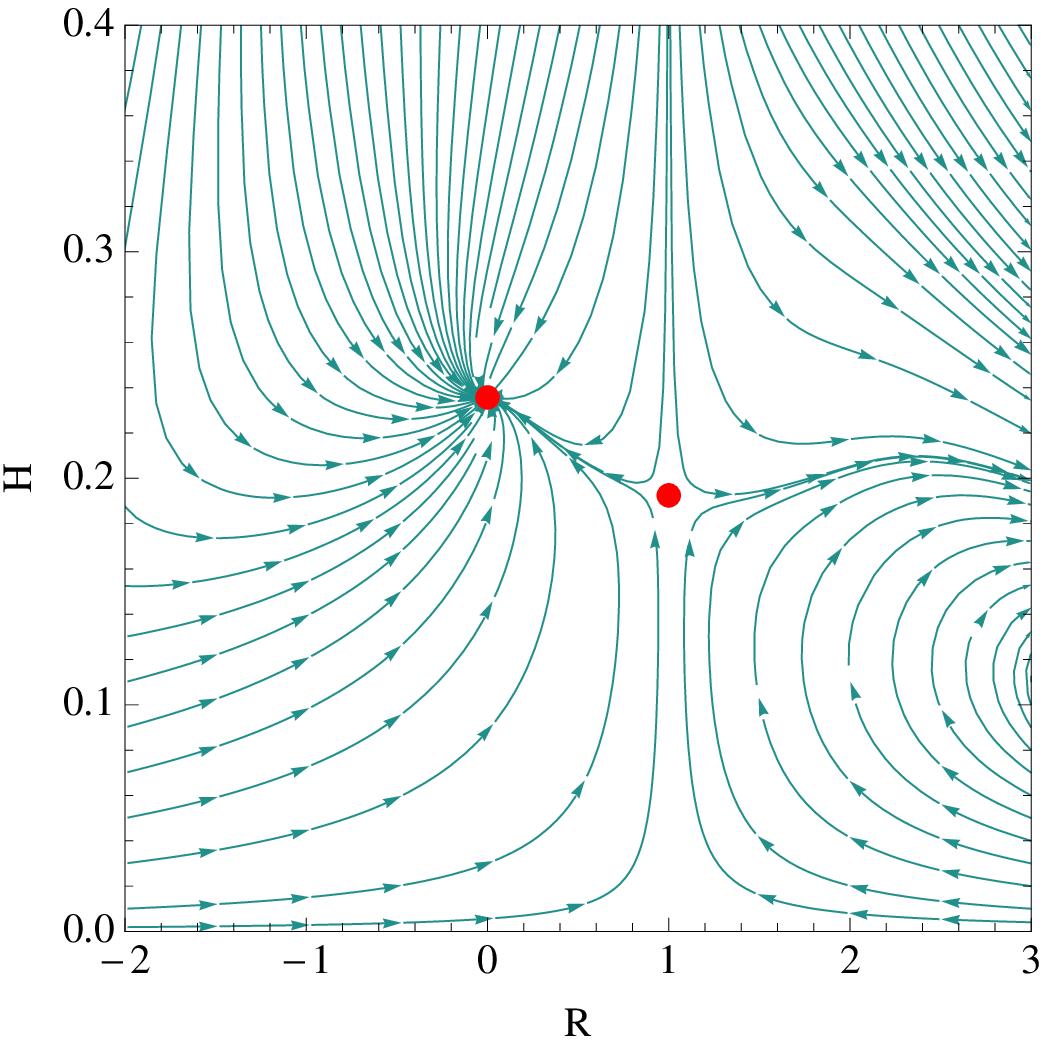}
\includegraphics[width=5.75cm]{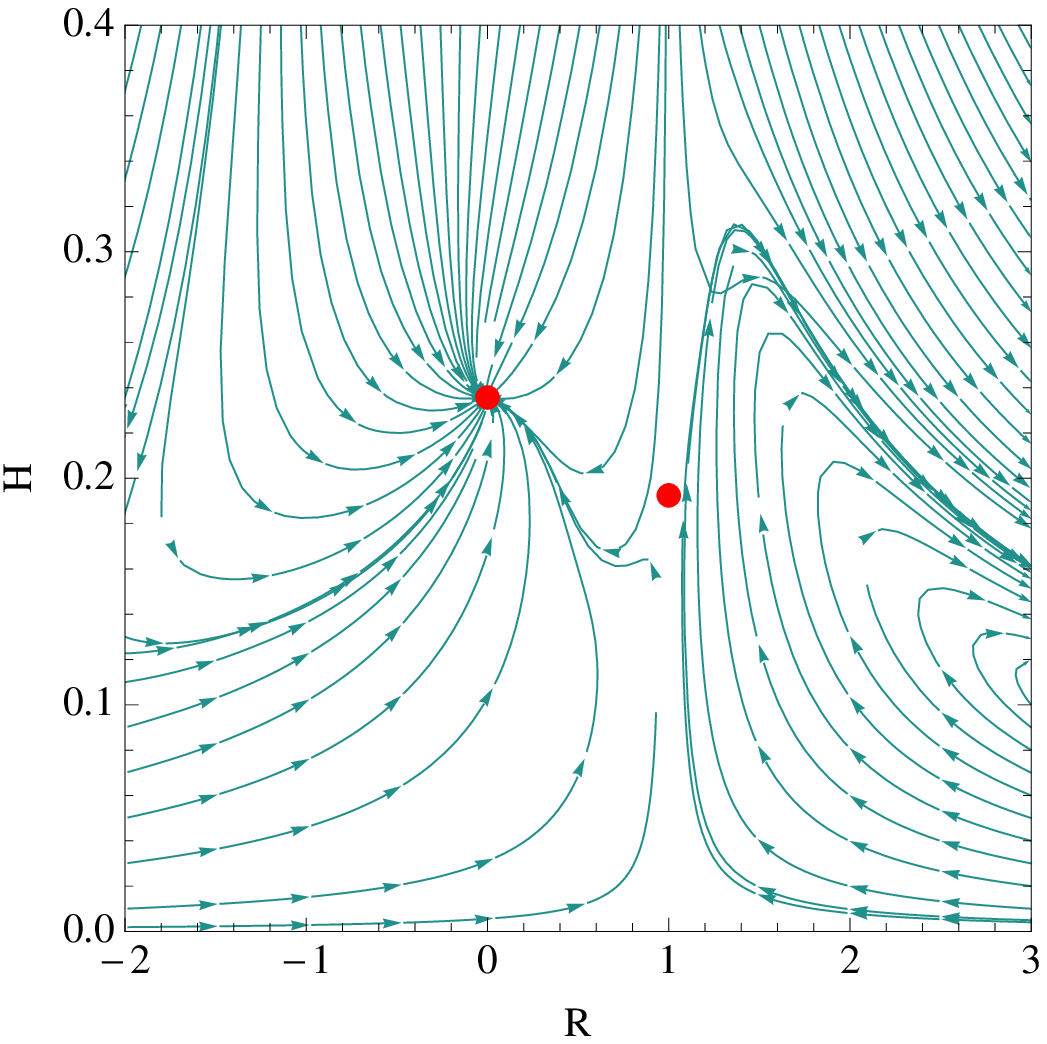}
\caption{We show the phase map for the autonomous system determining $H$ and $R$ for increasing values of $M^2$ (from left to right). We can see that the de Sitter critical point is an attractor, as discussed in the main text and that the critical point with $R=1$ is unstable. Moreover, neither the location nor the stability of these critical points depend on $M^2$; only the direction of the eigenvectors does. }
\label{phasemap}
\end{center}
\end{figure}

The first critical point corresponds to the isotropic ($R=0$) de Sitter solution driven by the cosmological constant. The second critical point is the electric Bianchi type I solution which is also a solution of the Einstein-Maxwell equations \cite{LeBlanc}. For this critical point we have $R=1$ so that $H_\perp=0$. This means that we have an exponential expansion along the longitudinal direction whereas the transverse ones remain frozen. However, this critical point is unphysical since the energy density of the vector field needs to be negative, as can be seen from the Friedmann constraint equation which leads to  $\rho_A=\dot A_z^2/(2a_\parallel^2) = -\rho_\Lambda $, that is negative for a positive cosmological constant.\footnote{Critical point II is physical if the cosmological constant is replaced by a perfect fluid with equation of state parameter $>1/3$ \cite{LeBlanc}.} It is interesting to note that the location of the critical points does not depend on $M^2$ so that the non-minimal coupling does not play any role. The linearized system for each critical point reads
\begin{eqnarray}
\frac{d}{dt}\left(\begin{array}{c}
\delta R \\ \delta  H\end{array}\right)  &=& M_{I,II} \left(
\begin{array}{c}
\delta R \\ \delta H\end{array}\right) 
\end{eqnarray}
with the corresponding matrices given by
\begin{eqnarray}
M_I &=&\left(
\begin{array}{cc}
-\sqrt{\frac{3\rho_\Lambda}{M^2M_p^2}} \;\;\;\;&-4\frac{1+\frac{2\rho_\Lambda}{M^2M_p^2}}{1+\frac{4\rho_\Lambda}{M^2M_p^2}}\\&\\ \;\;\;\;0 & -4\sqrt{\frac{\rho_\Lambda}{3M_p^2}}
\end{array}
\right);\;\;\;\;\;\;
M_{II}~=\left(
\begin{array}{cc}
\sqrt{\frac{2\rho_\Lambda}{M_p^2}} & \;\;\;\;\;0\;\;\;\\&\\ -\frac{4}{9}(1-\frac{4\rho_\Lambda}{M^2M_p^2})\frac{\rho_\Lambda}{M_p^2} &\;\;\;\;-2\sqrt{\frac{2\rho_\Lambda}{M_p^2}}
\end{array}
\right) \,.
\end{eqnarray}
These matrices show that the critical point I (corresponding to the de Sitter universe) is stable whereas the point II corresponds to a saddle point. Interestingly the stability of the critical points do not depend on $M^2$ either; only the direction of the eigenvectors as well as one of the eigenvalues depend on $M$. One of the eigenvectors of $\rm M_{II}$ is $(0,1)$. This direction is in fact a separatrix, as we see in Fig. \ref{phasemap}, so that if the initial state of the universe has $R<1$, i.e., 
$H_\perp>0$, it will end up in a de Sitter phase. This simply reflects the fact that the region on the right of the separatrix is unphysical because the energy density of the vector field is negative in this region  (as we discussed concerning the critical point II above) so that the system cannot evolve towards there coming from $R<1$. Thus, the only physical attractor solution is the pure isotropic de Sitter universe. This is in fact the expected result according to the well-known fact that shear decays in the presence of a cosmological constant ~\cite{Barrow:1995fn}.  We should mention that we have restricted the analysis to the spatially flat case (i.e., pure Bianchi I metric) and, in such a case,  our results indicate that the stability of the de Sitter universe is not spoiled by the presence of the non-minimal interaction. Nevertheless, we do not expect that such a conclusion would be modified by the presence of spatial curvature (like in more general Bianchi metrics), although additional critical points might appear in the phase map.

\section{Hamiltonian stability analysis \label{hamiltonian}}
Recently, there have been an increasing efforts to understand the classical and quantum (in)stability of vector theories \cite{vectorInstability,viability}.  Since it respects the gauge symmetry and yields second order field equations, it was mentioned in \cite{EspositoFarese:2009aj} that the Horndeski interaction could be viable, but a thorough investigation of this is to the best of our knowledge not previously presented in the literature.
Thus, our purpose in this section is to consider some relevant spacetime backgrounds  and derive the conditions for the absence of ghosts and Laplacian instabilities of our vector Langrangian on these backgrounds. Let us recall that a ghost is a field with negative kinetic energy while a Laplacian instability implies that we (formally) have negative squared propagation speed  for high enough frequencies. Together these conditions imply a lower bound free Hamiltonian. Note that the $U(1)$ gauge invariance of the theory will prevent the generation of an effective mass of the vector field so that no tachyonic instabilities associated with negative  square masses will appear. Let us also repeat that in this analysis the vector field is considered as subdominant, a test field.

\subsection{de-Sitter background \label{section:deSitter}}
Let us first study  the theory in the maximally symmetric de Sitter spacetime
\be
ds^2 = \frac{1}{(H\eta)^{2}}\Big(d\eta^2-dx^2-dy^2-dz^2\Big),
\ee
where $H$ is the constant Hubble expansion rate and $\eta<0$ for the expanding solution; $\eta =0$ is the Cauchy  horizon. In this case, the Riemann tensor can be fully written in terms of the metric as 
\begin{equation}
R_{\alpha\beta\gamma\delta}=H^2(g_{\alpha\delta}g_{\beta\gamma}-g_{\alpha\gamma}g_{\beta\delta}).
\end{equation}
Moreover, in this background,  the dual tensor reduces to $L^{\alpha\beta\gamma\delta}=2R^{\alpha\beta\gamma\delta}$ so that it is also given in terms of the metric. Using these relations we obtain the following expression for the effective action of the vector field in de Sitter
\begin{equation}
S=\int d^4 x\sqrt{-g}\left[-\frac{1}{4}\left(1+\frac{4 H^2}{M^2}\right)F_{\mu\nu}F^{\mu\nu}\right].
\end{equation}
Since the total Lagrangian is proportional to the usual Maxwell Lagrangian, it follows immediately that the theory is ghost-free and has a  Hamiltonian which is bounded from below if and only if 
\begin{equation}
1+\frac{4 H^2}{M^2}>0.\label{badcond}
\end{equation}
Next, as a warm-up for the next section, we consider the equations of motion. Since the background metric is homogeneous, we decompose the vector field in its Fourier modes with respect to the spatial coordinates
\be
\A_\mu=\int \frac{d^3k}{(2\pi)^{3/2}} \A_{\mu,\K}\;(\eta)e^{i\K\cdot\vec{x}},
\label{fourierA}
\ee
in terms of which the equations of motion adopt the following form:
\begin{eqnarray}
\A_{0,\K}\K+i\vec{\A}'_{\parallel,\K}=0 \label{eqofmotion:desitter1},\\
\vec{\A}_{\perp,\K}''+k^2\vec{\A}_{\perp,\K }=0,\\
\vec{\A}''_{\parallel,\K}-i \A'_{0,\K}\K=0, \label{eqofmotion:desitter3}
\end{eqnarray}
where $'=d/d\eta$, $\vec{\A}_{\parallel,\K}= \vec{k}(\vec{k}\cdot \vec{\A}_{\K})/k^2$ is the longitudal part of the vector field and $\vec{\A}_{\perp,\K}=\vec{\A}_{\K}-\vec{\A}_{\parallel,\K}$ is the transverse part satisfying $\vec{k}\cdot\vec{\A}_{\perp,\K}=0$. Since the Maxwell term is conformally invariant, the modes are not coupled to the expansion.\footnote{Incidentally, the conformal invariance of the full vector field sector including the non-minimal coupling is not broken in de Sitter. This is no longer true in more general FLRW backgrounds though.} Also, we have not made any gauge choice yet, so we can set $\vec{\A}'_{\parallel,\K}=0$ and, then, $\A_{0,\K}=0$  as well by virtue of the above equations. On the other hand, the physical transverse modes are given by the usual plane wave solutions so that the solution for the vector field takes the following form: 
\be
\vec{\A}=\int\frac{d^3k}{(2\pi)^{3/2}}\frac{1}{\sqrt{2k}}\sum_{\lambda=1,2}\vec{\epsilon}^{\;\lambda}_{\vec{k}} a^{\lambda}_{\vec k} e^{-(ik\eta-\K\cdot\vec{x})}+c.c,
\ee
where $\vec{\epsilon}_{\vec{k}}^{\; \lambda}$ are the normalized polarization vectors satisfying $\vec{\epsilon}_{\vec{k}}^{\; \lambda}\cdot\vec{k}=0$ and $\vec{\epsilon}_{\vec{k}}^{\; \lambda}\cdot\vec{\epsilon}_{\vec{k}}^{\; \lambda'}=\delta^{\lambda\lambda'}$. As expected, the propagation speed for the transverse modes is 1. This simply reflects the conformal invariance of the vector field action in the de Sitter background, so that the solutions look like in a Minkowski background.
When we insert these plane wave solutions into the corresponding expression for the energy-density given in (\ref{Tmunu}), we obtain\footnote{When expanding the squared electric or magnetic field ($E^2$ or $B^2$) in the Fourier modes of the vector potential, there appear terms of the type  $a^{\lambda}_{\vec k}a^{*\lambda}_{-\vec k} e^{-2ik\eta}$. These terms  exactly cancel out in the energy-momentum tensor in the case of ordinary Maxwell theory. However, the presence of the non-minimal interaction spoils such a cancellation and indeed terms of the type  $(H/M)^2a^\lambda_{\vec k}a^{*\lambda}_{-\vec k} e^{-2ik\eta}$ are present in the energy momentum tensor.  Here, we have disregarded these terms because, due to the harmonic time dependence, they become insignificant when integrating over several periods. Thus, the given energy density should actually be interpreted as the energy density averaged over several periods.}
\begin{equation}
\rho=\left(1+\frac{2H^2}{M^2}\right)\int \frac{d^3k}{a^4}\vert\K\vert \sum_{\lambda=1,2}\vert a^{\lambda}_{\vec k}a^{*\lambda}_{\vec{k}}\vert^2.
\end{equation}
We note that the energy density is positive if 
\begin{equation}
\left(1+\frac{2H^2}{M^2}\right)>0,
\end{equation}
which is a weaker condition than (\ref{badcond}). The energy density  will therefore always be positive when the Hamiltonian is bounded from below.  Finally, notice that the above expression for the energy density can be recast into the usual expression in flat spacetime by a rescaling of the Fourier amplitudes, as it was known from the aforementioned conformal invariance of the Horndeski action in pure de Sitter.

\subsection{Stability in a flat FLRW background \label{flatFLRW}}
In the following we will study the stability of the Horndeski term in a flat FLRW background
\be
ds^2 = a(\eta)^2 (d\eta^2-d\vec{x}^2),
\ee
and consider a general vector potential $\A=\A_\alpha(\eta,x,y,z) \diff x^\alpha$.

We introduce the electric and magnetic components, relative to an observer with four velocity $u^\mu$, of the Faraday tensor defined by~\cite{magnet}
\begin{eqnarray}
E_\mu&\equiv& F_{\mu\nu}u^\nu,\\
B_\mu&\equiv &\Ft_{\mu\nu} u^\nu=\frac12\epsilon_{\mu \nu \alpha\beta} F^{\alpha\beta}u^\nu,
\end{eqnarray}
where $\epsilon_{\mu \nu \alpha\beta}=\epsilon_{[\mu \nu \alpha\beta]}$ is the volume element with sign convention $\epsilon_{0123}=\sqrt{-g}$.
In terms of these vectors, the Faraday tensor can be expressed as
\be
F_{\mu\nu}=2E_{[\mu}u_{\nu]}+\epsilon_{\mu\nu \alpha\beta}B^\beta u^\alpha.
\ee 
We consider the congruence of comoving observers $u^{\mu}=(a^{-1},0,0,0)$ and express the action in terms of these variables: 
\be
S=-\frac12\int\diff^3x\diff\eta \; a^4\!\left[\left(1+\frac{4\HH^2}{a^2M^2}\right) E_\mu E^\mu-\left(1-\frac{4q\HH^2}{a^2M^2}\right) B_\mu B^\mu \right],
\label{EBactionFLRW}
\ee
where $q=-\HH'/\HH^2$ is the deceleration parameter. From this expression we can straightforwardly read off the Hamiltonian stability conditions. This becomes more apparent when we write the action in terms of the vector potential,
\be
\begin{split}
S&=\frac12\int \diff^3x\diff \eta \left[\left(1+\frac{4\HH^2}{a^2M^2}\right)(\vec{\A}_{\perp}')^2-\left(1-\frac{4q\HH^2}{a^2M^2}\right)\big(\vec{\nabla}\times\vec{\A}_{\perp}\big)^2\right] \\
&=\frac12\int \diff^3k\diff \eta \left[\left(1+\frac{4\HH^2}{a^2M^2}\right)(\vec{\A}_{\perp,\vec{k}}')^2+\left(1-\frac{4q\HH^2}{a^2M^2}\right)k^2\big(\vec{\A}_{\perp,\vec{k}}\big)^2\right],
\end{split}
\label{actionFLRW}
\ee
where the cross denotes the ordinary cross product between the three vectors defined by $\vec\nabla=(\partial_x,\partial_y,\partial_z)$ and $\A_\mu=(0,\vec{\A}_\perp)$. Here we have again fixed the gauge by requiring that the longitudinal component vanishes, $\nabla \cdot \vec\A=0$ which implies $\A_0=0$, as we shall confirm from the equations of motion below, and hence $\vec{\A}_\perp$ represents the physical transverse modes.  For the second equal sign of (\ref{actionFLRW}) we have expanded the action in Fourier coefficients, choosing the normalization such that the Fourier transform is unitary. Now we can directly read off the stability conditions. 
Note that the Hamiltonian stability is determined by the factors in round brackets preceding the $E_\mu E^\mu$ and $B_\mu B^\mu$ terms in (\ref{EBactionFLRW}). We shall use this in the following sections as well. 

First, to have positive kinetic terms we must require
\be
1+\frac{4H^2}{M^2} > 0,
\label{ghostfreeFLRW}
\ee
which is the condition for absence of ghosts\footnote{Curiously, this condition coincides exactly with the condition for avoidance of the singularity (where the deceleration parameter diverges within a finite proper time) identified in \cite{Barrow:2012ay}.}. Next, as we shall confirm from the equations of motion below, the propagation speed, $c_s$, in the high frequency (short wavelength) regime, is simply the ratio of the factors in round brackets,
\be
c_s^2=\frac{1-\frac{4qH^2}{M^2}}{1+\frac{4H^2}{M^2}} = 1-\frac{\frac{4H^2}{M^2}(1+q)}{1+\frac{4H^2}{M^2}}.
\label{speed:flrw}
\ee
Using the geometric optics approximation, Ref.~\cite{Teyssandier} has calculated the propagation speed for a general non-minimal coupling linear in the curvature and quadratic in  the Faraday tensor. Specializing to the Horndeski action, the propagation speed obtained in~\cite{Teyssandier} agrees with our equation (\ref{speed:flrw}).\footnote{In the case of Horndeski's Lagrangian, the three non-minimal coupling parameters in \cite{Teyssandier} are related to $M^2$ as follows: $2/M^2=\xi=-\eta/2=\zeta$.}. In order to avoid the Laplacian instability we must require
\be
c_s^2\ge0 \,.
\label{laplacianinstabilityFLRW}
\ee
This ensures that the equation of motion for $\vec{\A}_\perp$ takes the form of a wave equation and, thus, no exponentially growing solutions exist.  The combined conditions for the avoidance of ghosts (\ref{ghostfreeFLRW}) and Laplacian instabilities (\ref{laplacianinstabilityFLRW}) imply a Hamiltonian which is bounded from below,
\be
H=\frac12\int \diff^3x \left[\left(1+\frac{4H^2}{M^2}\right)\vert\vec{\A}_\perp'\vert^2+\left(1-\frac{4qH^2}{M^2}\right)\big(\vec{\nabla}\times\vec{\A}_\perp\big)^2\right]. 
\ee

\begin{figure}[htbp]
\begin{center}
\includegraphics[width=0.5\textwidth]{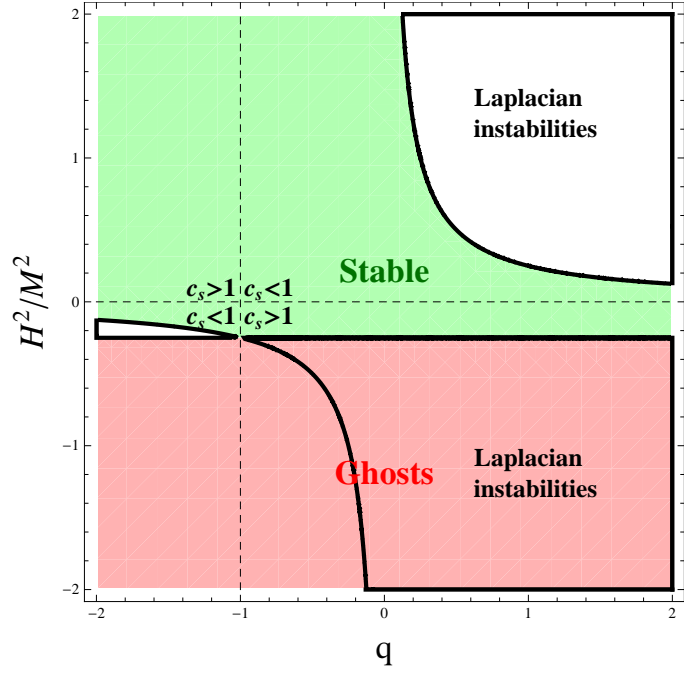}
\end{center}
\caption{The stable region for FLRW cosmologies is indicated by green. The red region is plagued by ghosts, while the region enveloped by the thick black lines are plagued by Laplacian instabilities. The dashed lines separate the super-luminous and sub-luminous regions. }
\label{fig:flrw}
\end{figure}

In Fig. \ref{fig:flrw} we have indicated the regions plagued by instabilities using $q$ and $H^2/M^2$ as independent variables. We note the existence of stable regions with both  positive and negative $M^2$.  The horizontal dashed line at $H^2/M^2=0$ represents conventional Maxwell theory in a FLRW background. We note that the Maxwell theory has a stable neighborhood. The vertical dashed line at $q=-1$ represents de-Sitter spacetime which is stable for $4H^2/M^2>-1$. For $4H^2/M^2<-1$ the Hamiltonian is not bounded from below.  Furthermore, as indicated in the figure, these dashed lines also separate regions with super-luminous and sub-luminous propagation speed (this does of course not apply to the regions with Laplacian instabilities where $c_s^2$ is negative). 
From (\ref{speed:flrw}) it follows that
\be
c_s^2\le1 \Rightarrow -\frac{q}{M^2} \le \frac{1}{M^2}.
\ee
Hence, the condition for propagation speed less than or equal to $1$ is $q\ge -1$ for $M^2>0$, and $q\le -1$ for $M^2<0$. In other words, for positive $M^2$ the propagation speed is less than or equal to $1$ if the null energy condition holds and greater to one if it is violated. Conversely, for negative $M^2$ the propagation speed is smaller than $1$ only if the null energy condition is violated.   

It is interesting to note that, in a power law expanding universe, the Laplacian instability will eventually cease and the vector field will enter into the stable region (for positive $M^2$). This is so because, for a power law expansion, the deceleration parameter $q$ is constant, so that the system evolves along a vertical line in Fig \ref{fig:flrw} and towards the $x$-axis. Thus, it could start in the region of Laplacian instabilities but, at some point, it will cross to the stable region. This is easy to understand because the impact of the Horndeski interaction decreases in an expanding universe so that our action renders the Maxwell theory at late times, which is of course stable.

Next, we study the equations of motion.  First, without any gauge choice, following a procedure similar as for the de-Sitter spacetime in section \ref{section:deSitter}, we decompose the $\A_\mu$ in its Fourier modes so that the equations of motion adopt the following form:
\begin{eqnarray}
\A_{0,\K}\K+i\vec{\A'}_{\parallel,\K}=0, \label{eqofmotionFLRW1}\\
\left( 1+\frac{4\HH^2}{a^2M^2} \right)\vec{\A}_{\perp,\K}'' -\frac{8\HH^2}{a^2M^2}(1+q)\HH\vec{\A}_{\perp,\K}'+k^2 \left(1-\frac{4q\HH^2}{a^2M^2} \right)\vec{\A}_{\perp,\K }=0, \label{eqofmotionFLRW2}\\
\left( 1+\frac{4\HH^2}{a^2M^2} \right)\vec{\A}_{\parallel,\K}'' -\frac{8\HH^3}{a^2M^2}(1+q)\vec{\A}_{\parallel,\K}'-i\K \left[ \left( 1+\frac{4\HH^2}{a^2M^2} \right) \A'_{0,\K}-\frac{8\HH^3}{a^2M^2}(1+q)\A_{0,\K} \right]=0.\label{eqofmotionFLRW3}
\end{eqnarray}
We note that these equations reduce to (\ref{eqofmotion:desitter1})-(\ref{eqofmotion:desitter3}) for the deceleration parameter of de Sitter, $q=-1$.  We now  fix the gauge choice by setting $\vec{\A}_{\parallel,\K}=0$ in which case (\ref{eqofmotionFLRW1}) implies $\A_0=0$ as claimed above.  In this gauge (\ref{eqofmotionFLRW3}) becomes redundant and the dynamics of the vector potential $\vec{\A}_{\K}=\vec{\A}_{\perp,\K}$ is given by (\ref{eqofmotionFLRW2}) alone. From this we now find under which conditions the propagation speed, $c_s$, from the action, is valid.  For convenience, we first write (\ref{eqofmotionFLRW2}) on the form 
\be
\alpha \vec{\A}_{\perp,\K}'' +\beta \HH \vec{\A}_{\perp,\K}'+ k^2  \gamma \vec{\A}_{\perp,\K }=0,
\label{WKBeq}
\ee
with obvious definitions for the coefficients $\alpha$, $\beta$ and $\gamma$. In the high-frequency regime the time variation of $\HH$ and $q$ are negligible over time scales corresponding to several periods for $\vec{\A}_{\perp,\K}$.  This is a good approximation for wavelengths much smaller than the horizon, i.e., $\HH/k\ll1$.  We can then solve~(\ref{WKBeq}) in the WKB approximation which yields
\be
\vec{\A}_{\perp,\K} \propto \exp{\left[-\int d\eta\HH \left( \frac{\beta}{2\alpha} \pm \sqrt{\frac{\beta^2}{4\alpha^2}-\frac{k^2}{\HH^2}\frac{\gamma}{\alpha}} \right) \right]}.
\label{ligning}
\ee    
From (\ref{speed:flrw}), apparently we have $c_s\rightarrow\infty$ when $4H^2/M^2\rightarrow-1^+$.This corresponds to the limit $\alpha\rightarrow0^+$ in which case the solutions are
\be
\vec{\A}_{\perp,\K} \propto \exp{\left[-\int d\eta\frac{\HH\beta}{2\alpha} \left( 1\pm\left(1 -\frac{2k^2\alpha\gamma}{\HH^2\beta^2}\right) \right) \right]} \, .
\label{ligning3}
\ee    
Using that $\gamma/\beta=1/2$ in the considered limit, the solution with the minus sign becomes
\be
\vec{\A}_{\perp,\K} \propto \exp{\left[-\int d\eta\frac{k^2}{2\HH} \right]} \, .
\label{ligning2}
\ee    
This is the 'WKB approximation' of the solution to the diffusion equation into which~(\ref{WKBeq}) turns when $\alpha =0$. In this case $\vec{\A}_{\perp,\K}$ no longer satisfies a wave equation but a simple diffusion equation with only one solution. Therefore,  the solution with $+$ sign in (\ref{ligning}) vanishes when $\alpha\rightarrow0$.   

For $\alpha,\gamma>0$, which corresponds to the regime where the theory is stable, and
\be
\frac{\beta^2}{4\alpha^2}\ll\frac{k^2}{\HH^2}\frac{\gamma}{\alpha}
\label{condition}
\ee
we can write the WKB solutions in the form
\be
\vec{\A}_{\perp,\K} \propto \exp{\left[-\int \diff\eta\frac{\HH\beta}{2\alpha}\right]} \exp{\left[\pm ik\int \diff\eta c_s \right]},
\ee    
where $c_s=\sqrt{\gamma/\alpha}$ is the propagation speed which we have read off from the action. Since $k/\HH\gg1$ in the high frequency regime, the condition (\ref{condition}) always holds in the high-frequency regime unless $\alpha$ is close to zero.  In the critical case $\alpha \rightarrow 0$, we have $|\beta|\sim|\gamma|\sim 1$ and (\ref{condition}) can be rewritten 
\be
c_s \ll k^2/\HH^2.
\label{condition2}
\ee  
To summarize, we have shown that the propagation speed is finite and that $c_s^2=\gamma/\alpha$ is a good approximation in the high frequency regime when (\ref{condition2}) holds.  

To conclude this section, let us mention that the condition for the vector being subdominant can be written as
\be
\frac{1}{2} \left( \mathbf E^2 + \mathbf B^2\right) + 2 \frac{H^2}{M^2} \left( 3 \mathbf E^2 -2 \mathbf B^2\right) -  \frac{4 H}{M^2} \mathbf E \cdot (\boldsymbol{\nabla} \times \mathbf{B}) - \frac{2}{M^2} \boldsymbol{\nabla}\cdot \left[(\mathbf B \cdot \boldsymbol{\nabla}) \mathbf B  \right] \ll 3 M_p^2 H^2,
\ee
where $\mathbf E=\vec{\A}_{\perp}'/a^2$ and $\mathbf B=-\boldsymbol{\nabla}\times\vec{\A}_{\perp}/a$ are the electric and magnetic fields seen by a comoving observer and $\boldsymbol{\nabla} = (\partial_x, \partial_y, \partial_z)/a(t)$ is the Laplace operator in the corresponding local Lorentz frame. The left-hand side of the inequality is the energy density associated with the vector, while the right-hand side is the total energy density.  Note that the vector may well be subdominant even in the regime $H^2/M^2\gg1$, where it is dominated by the interaction terms. Thence, it is important to point out that the test-field assumption we have employed does not imply a bound on $M^2$ by itself.

\subsection{Stability in a curved FLRW background}
We briefly generalize the results of section \ref{flatFLRW} to a spatially curved FLRW background for completeness. In this case, the metric is given by
\be
\diff s^2 = a^2\left(\diff\eta^2 - \frac{\diff r^2}{1-Kr^2} - r^2\diff\theta^2 - r^2\sin^2\theta \diff\phi^2\right).
\ee
We write the action in the same form as before,
\be
S=-\frac12\int\diff^3x\diff\eta \; a^4\!\left[\left(1+\frac{4\HH^2}{a^2M^2}(1-\Omega_k)\right) E_\mu E^\mu-\left(1-\frac{4q\HH^2}{a^2M^2}\right) B_\mu B^\mu \right],
\label{EBactionCurvedFLRW}
\ee
where $\Omega_k=-K/\HH^2$ is the Hubble normalized curvature.  Then, following the approach explained in the previous section, we can easily obtain the conditions for absence of ghosts and Laplacian instabilities. In this case, such conditions generalize to
\be
1+\frac{4\HH^2}{a^2M^2}(1-\Omega_k) > 0
\ee
to guarantee the ghost freedom of the theory, whereas the condition
\be
c_s^2 = 1-\frac{\frac{4H^2}{M^2}(1+q-\Omega_k)}{1+\frac{4H^2}{M^2} (1-\Omega_k)} \ge 0,
\label{speed:curvedflrw}
\ee
prevents the theory from having Laplacian instabilities.  Again, our propagation speed agrees with \cite{Teyssandier}.

\subsection{Schwarzschild background}
In this section we will consider the gravitational field outside a spherical, uncharged and non-rotating object of mass $M_*$. This is described by the Schwarzschild metric
\begin{equation}
\diff s^2=\left(1-\frac{R_s}{r}\right)\diff t^2-\frac{\diff r^2}{\left(1-\frac{R_s}{r}\right)}-r^2\diff \Omega^2,
\end{equation}
where $R_s=2GM_*$ is the Schwarzschild radius (which determines the event horizon in the case of a black hole). Since this metric is Ricci flat, the action (\ref{action}) becomes
\be
S=\int \diff^4x\sqrt{-g}\left[-\frac14F_{\mu\nu}F^{\mu\nu} + \frac{1}{2M^2} R_{\mu\nu\alpha\beta} F^{\mu\nu} F^{\alpha\beta} \right],
\ee
which will allow us to compare the propagation speeds calculated below with the results of \cite{Drummond}.\footnote{In the case of Ricci flat spacetimes, the parameter $\xi^2$ used in \cite{Drummond} is related to our non-minimal coupling parameter by $\xi^2 = -2/M^2$.}  Introducing electric and magnetic fields for an observer at rest w.r.t the object,  we write the action in terms of the electric and magnetic fields as follows:
\be
S= -\frac{1}{2}\int \diff^4x \sqrt{-g} \Big[  \alpha E_r E^r + \beta\left( E_\theta E^\theta+ E_\phi E^\phi\right)- \alpha B_r B^r - \beta\left( B_\theta B^\theta+ B_\phi B^\phi\right) \Big],
\label{action:schwarzschild}
\ee
where
\begin{align}
\alpha&=1+ \frac{4R_s}{M^2r^3}, \\
\beta&=1-\frac{2R_s}{M^2r^3}. 
\end{align}

Restricting ourselves to the region outside the event horizon $(r>R_s)$, the condition for the absence of ghosts is $\alpha>0$ and $\beta > 0$, or  
\be
-1<\frac{4R_s}{M^2r^3}<2.
\label{ghostfreeSch}
\ee
We now want to read off the propagation speed and find the condition for the absence of Laplacian instabilities. We are interested in the propagation speed relative to the orthonormal frame
\be
\mathbf{e}_t = \left(1-\frac{R_s}{r}\right)^{-\frac{1}{2}}\partial_t, \quad \mathbf{e}_r = \left(1-\frac{R_s}{r}\right)^{\frac{1}{2}}\partial_r, \quad \mathbf{e}_\theta = r^{-1}\partial_\theta, \quad \mathbf{e}_\phi = (r \sin\theta)^{-1} \partial_\phi,
\ee
which represents the laboratory frame of a static observer. Note that, since the metric is diagonal, the  coefficients  $\alpha$ and $\beta$ do not change if we write the action in terms of the orthonormal frame components $\hat E_\mu$ and $\hat B_\mu$ (for instance $E_r E^r=\hat E_r  \hat E^r$, $B_r B^r=\hat B_r \hat B^r$ and so on for the other components). The action is therefore already in the desired form and, in principle, we can read off the propagation speed as in the sections above. Due to the broken spatial symmetries in the considered spacetime, however, we must use a little more care since the propagation speed will depend both on the direction of propagation and the direction of the polarization. First we consider a wave traveling in the radial direction.  Then we have $E_r=B_r=0$ and the propagation speed is
\be
c_r^2 = \frac{\beta}{\beta}=1.
\ee  
Thus, for a wave propagating in the radial direction, the propagation speed is not affected by the non-minimal interaction. As expected from the symmetries of the Schwarzschild spacetime, the velocity does not depend on the polarization since the electric and magnetic field will point in the tangential directions. For a wave traveling in the angular directions, however, the propagation speed depends on the polarization. For a wave traveling in the tangential direction  with polarization in the radial direction (for instance $E_\theta=E_\phi=B_\theta=B_r=0, E_r\neq 0, B_\phi\neq 0$) we read off the propagation speed
\be
c_{\Omega,1}^2 = \frac{\beta}{\alpha} = 1-\frac{\frac{6R_s}{M^2r^3}}{1+\frac{4R_s}{M^2r^3}}.
\ee  
For a wave traveling in the tangential direction  with polarization in the tangential direction (for instance $E_r=E_\theta=B_\theta=B_\phi=0,E_\phi\neq 0, B_r\neq 0$),  the propagation speed is
\be
c_{\Omega,2}^2 = \frac{\alpha}{\beta}=1+\frac{\frac{6R_s}{M^2r^3}}{1-\frac{2R_s}{M^2r^3}}.
\ee  
Our three propagation speeds agree with the results in \cite{Drummond} where an effective Lagrangian (which is equal to the Horndeski Lagrangian in the case of Ricci flat spacetimes) was considered in the context of vacuum polarization in ordinary QED. To lowest order in curvature, we have $c_{\Omega,1}^2=1-6R_s/(M^2r^3)$ and $c_{\Omega,2}^2=1+6R_s/(M^2r^3)$. Therefore, it is clear that super-luminal propagation speed is an unavoidable consequence of the non-minimal interaction in Schwarzschild spacetime. The condition to avoid Laplacian instabilities, i.e. to have real propagation speeds for all three modes, can be summarized as 
\be 
-1\le \frac{4R_s}{M^2r^3} \le 2.
\label{lapfreeSch}
\ee 
Apart from the boundary,  where the sound speed actually diverges and the wave equation becomes a Laplace equation, this is identical to the condition for the absence of ghosts (\ref{ghostfreeSch}).

\begin{figure}[htbp]
\begin{center}
\includegraphics[width=0.5\textwidth]{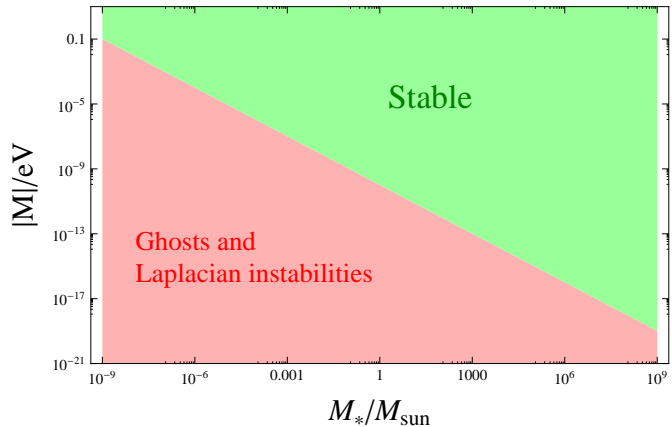}
\end{center}
\caption{The stable region in green and the region plagued by ghosts and Laplacian instabilities in red, obtained from equation (\ref{ghostfreeSch1b}). $M^2$ is the non-minimal coupling constant, while $M_*$ is the mass of the star and $M_\text{sun}$ the solar mass.}
\label{fig:schwarzschild}
\end{figure}

The entire exterior region ($r>R_s$) is therefore free of ghosts and Laplacian instabilities if
\be
-\frac{1}{4}R_s^2 < \frac{1}{M^2} < \frac{1}{2} R_s^2,
\label{ghostfreeSch1}
\ee
which can be rewritten
\be
-\left( \frac{M_*}{M_\text{sun}} \right)^2 < 1.8\times 10^{-20} \left(\frac{eV}{M}\right)^2 < 2\left( \frac{M_*}{M_\text{sun}} \right)^2, \\
\ee
or roughly
\be
\frac{|M|}{eV} \gtrsim 10^{-10} \frac{M_\text{sun}}{M_*},
\label{ghostfreeSch1b}
\ee
where $M_*$ is the mass of the black hole. This condition obviously also applies to the case of the exterior region of a star for which the Schwarzschild radius is inside the objet. Fig. \ref{fig:schwarzschild} displays the stable region obtained from (\ref{ghostfreeSch1b}). We note that a smaller black hole mass implies a stronger constraint on the non-minimal coupling parameter $M^2$, which is a consequence of the fact that the scalar curvature at the event horizon is proportional to $1/R_s^2$.  
 
From the action (\ref{action:schwarzschild}) we note that the ratio between the non-minimal and minimal part of the Lagrangian density is of order $\mathcal{L}_H / \mathcal{L}_M \sim 2R_s/(M^2r^3)$. The Horndeski interaction is most important close to the event horizon where we have $(\mathcal{L}_H / \mathcal{L}_M)_{r=R_s} \sim 2/(M^2R_s^2)$.  The stability condition (\ref{ghostfreeSch1}) can then be rewritten
\be
-1/2 \lesssim \left(\frac{\mathcal{L}_H}{\mathcal{L}_M}\right)_{r=R_s} \lesssim 1,
\label{ghostfreeSch2}
\ee
which means that the ratio $(\mathcal{L}_H/\mathcal{L}_M)$ can  be at most  of order unity at the event horizon, and further away it will be even more suppressed.  For instance already at $r=10R_s$ we have $|\mathcal{L}_H/\mathcal{L}_M|\lesssim 10^{-3}$ which means that the non-minimal part of the interaction is already negligible. 

To summarize, we have shown that having a stable exterior region of a Schwarzschild black hole implies that the non-minimal part of the interaction nowhere dominates over the minimal one.  Any possible observable effect of the theory from strong gravitational environments must therefore stem from a tiny region close to the event horizon where the non-minimal part of the interaction may match the minimal one, while further out it quickly falls off and becomes negligible. Turning this the other way around; regions where the Horndeski terms dominate the interaction do have ghosts and Laplacian instabilities.  We conclude that although there is a theoretical possibility for the non-minimal interaction to play some role near the event horizon of a black hole, any possible astrophysical signature must arguably be very weak and it is doubtful whether it could ever be detected.

\subsection{Reissner-Nordstrom-de Sitter background}
In this section, we generalize the results of the Schwarzschild metric to the case of of a Reissner-Nordstrom-de Sitter metric corresponding to a black hole of mass $M_*$,  electric charge $Q$ and in the presence of a cosmological constant $\Lambda$. We shall give the results here for completeness, although black holes with an appreciable amount of electric charge are likely not to exist in nature. The line element for this metric reads
\be
\diff s^2=A(r)\diff t^2-A^{-1}(r)\diff r^2-r^2\diff\Omega^2,
\ee
with 
\be
A(r)=1-\frac{R_s}{r}+\frac{Q^2}{r^2}-\frac13\Lambda r^2.
\ee
This spacetime has three horizons\footnote{The three horizons correspond to the zeros of the function $A(r)$. Since the equation $A(r)=0$ leads to a quartic polynomial equation, there will be 4 solutions. However, the absence of cubic term makes that one of the solutions is always negative and, hence, non-physical.} corresponding to an inner Cauchy horizon, the black hole event horizon and, finally, the most external de Sitter horizon. It is important to keep in mind that, if the non-minimal coupling in the action is appreciable such that its backreaction on the spacetime geometry cannot be neglected, the above metric is no longer a solution in the case where the charge $Q$ corresponds to the same $U(1)$ gauge field as the one appearing in the Horndeski coupling\footnote{See \cite{MuellerHoissen:1988bp} for a discussion on static and spherically symmetric solutions to Einstein-Maxwell equations with the additional Horndeski interaction term}. Thus, either this coupling term has to be subdominant or we have to consider two different $U(1)$ gauge fields, one related to the charge $Q$ with a simple Maxwell action and one which is unrelated to $Q$ but which has the non-minimal Horndeski coupling. In the subsequent analysis we shall adopt the second approach as we are interested in studying the stability of the Horndeski lagrangian in the given spacetime background.

As in the previous section, we can write the effective action in this background
\be
S= -\frac{1}{2}\int \diff^4x \sqrt{-g} \Big[  \alpha_E E_r E^r + \beta_E\left( E_\theta E^\theta+ E_\phi E^\phi\right)- \alpha_B B_r B^r - \beta_B\left( B_\theta B^\theta+ B_\phi B^\phi\right) \Big],
\label{action:RNdS}
\ee
where now the coefficients generalize to
\begin{align}
\alpha_E&=1+ \frac{4R_s}{M^2r^3}-\frac{4Q^2}{M^2r^4}+\frac{4\Lambda}{3M^2}, \\
\alpha_B&=1+ \frac{4R_s}{M^2r^3}-\frac{12Q^2}{M^2r^4}+\frac{4\Lambda}{3M^2},\\
\beta_E&=1-\frac{2R_s}{M^2r^3}+\frac{4Q^2}{M^2r^4}+\frac{4\Lambda}{3M^2},\\
\beta_B&=1- \frac{2R_s}{M^2r^3}+\frac{4Q^2}{M^2r^4}+\frac{4\Lambda}{3M^2}.
\end{align}
The requirement for the absence of ghost instabilities is then $\alpha_E >0$ and $ \beta_E >0$ in order not to have ghost-like modes propagating along the radial and angular directions respectively. We can see that the cosmological constant contribution always plays in favor of the absence of ghost instabilities for $M^2>0$. This is not at all surprising since at large enough distances as compared to the black hole event horizon, the spacetime structure is dominated by the cosmological constant contribution so that we essentially have a de Sitter universe. In such a region, the conditions for the absence of ghost instabilities of the vector field are fulfilled for $M^2>-\frac 43\Lambda$, although $M^2>0$ is a sufficient condition. In fact, we can see that, in this case, all the coefficients in the action become the same and equal to $1+\frac{4\Lambda}{3M^2}$, which exactly coincides with our previous findings in section \ref{section:deSitter} for a purely de Sitter background (where $H^2=\Lambda/3$). 

\begin{figure}[htbp]
\begin{center}
\includegraphics[width=0.4\textwidth]{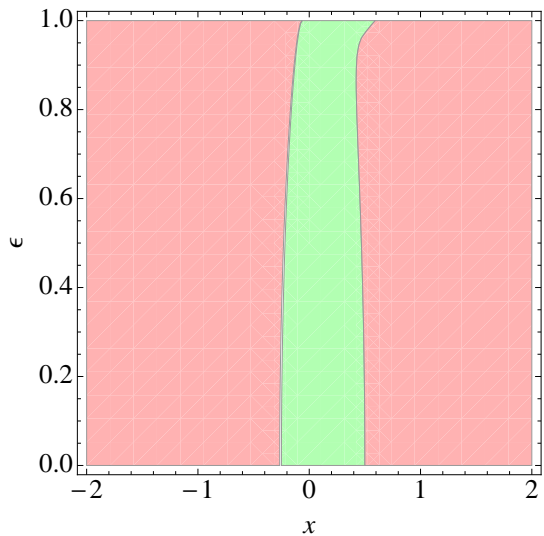}
\end{center}
\caption{In this plot we show the stable region (in green)  of  the parameter space $\Big(x=(MR_s)^{-2}, \epsilon= 2Q/R_s,\Big)$ for the Reisner-Nordstr\"om spacetime. Notice that all the equations depend on $\epsilon^2$, so that the region with $\epsilon<0$ is the same as the one shown here. We can see that for $\epsilon=0$ we recover the Schwarzschild bounds and that the stability conditions barely depend on the black hole electric charge.}
\label{RN}
\end{figure}

In the region near the event horizon of the black hole we can neglect the cosmological constant term\footnote{We are assuming here that the de Sitter horizon is far away from the black hole event horizon so that near such a horizon we can safely neglect the cosmological constant contribution. In other words, we are assuming that there exists a region in which the metric can be well approximated by the pure Reissner-Nordstr\"om. This is in fact a reasonable assumption given that the {\it observed} value of the cosmological constant is very small $\rho_\Lambda\simeq (10^{-3}{\rm eV})^4$. Thus, our assumption holds for black holes with masses $M_*\ll10^{50} M_\odot$, which is a quite safe assumption for our universe.}. Then, the black hole event horizon is approximately given by $r_{\rm eh}\simeq (R_s+\sqrt{R_s^2-4Q^2})/2$ (up to corrections ${\mathcal O}(\sqrt{\Lambda}$). Notice that it is necessary to have $R_s\geq2|Q|$  in order to dress up the singularity at the origin so we shall impose this condition in our analysis. In this region, the conditions for the avoidance of ghost instabilities reduce to 
\begin{eqnarray}
&&\alpha_E\simeq 1+\frac{32 x}{\rr^3\left(1+\sqrt{1-\epsilon^2}\right)^3}\left[1-\frac{\epsilon^2 }{2\rr\left(1+\sqrt{1-4\epsilon^2}\right)}\right] >0,\\
&&\beta_E\simeq 1-\frac{16 x}{\rr^3\left(1+\sqrt{1-\epsilon^2}\right)^3}\left[1-\frac{\epsilon^2 }{\rr\left(1+\sqrt{1-\epsilon^2}\right)}\right] >0,
\end{eqnarray}
where we have introduced a radial coordinate normalized to the event horizon $\rr\equiv r/r_{\rm eh}$, the ratio of the black hole charge to its mass $\epsilon\equiv 2Q/R_s=Q/(GM_*)$ (that must satisfy $\epsilon\leq1)$ and $x\equiv (M R_s)^{-2}$, which essentially controls the magnitude of the Horndeski term as compared to the Maxwell term. The exact stability conditions are shown in Fig. \ref{RN}. In order to obtain that plot we have minimized the above expressions with respect to $\rr$ restricted to the region outside the event horizon $\rr>1$ in order to guarantee stability at all scales. In that plot we see that such conditions depend very mildly on $\epsilon$. This is easy to understand from the above expressions since $\epsilon$ is restricted to $\epsilon\leq 1$. Then, we can Taylor expand those expressions and see that the correction due to the black hole having a non-vanishing electric charge is ${\mathcal O}(\epsilon^2)$ so that the absence of ghosts is a stable feature against the introduction of electric charge for the black hole.

Concerning the presence of laplacian instabilities, we need to consider modes propagating along radial and angular directions with respect to the corresponding orthonormal frame,  as we did in the previous section for the Schwarzschild spacetime. For a wave traveling along the radial direction, the propagation speed is given by
\be
c_{r}^2=\frac{\beta_B}{\beta_E}=1\, .
 \ee
Thus, as we obtained for the Schwarzschild metric, the modes propagating in radial directions never become unstable and remain unaffected by the presence of the Horndeski interaction. On the other hand, for the angular modes  we find the following propagation speeds:
 \begin{eqnarray}
c_{\Omega,1}^2 &=&\frac{\beta_B}{\alpha_E}=1-\frac{\frac{6R_s}{M^2r^3}-\frac{8Q^2}{M^2r^4}}{1+\frac{4R_s}{M^2r^3}-\frac{4Q^2}{M^2r^4}+\frac{4\Lambda}{3M^2}}\, ,\\\nonumber\\
c_{\Omega,2}^2 &=&\frac{\alpha_B}{\beta_E}=1+\frac{\frac{6R_s}{M^2r^3}-\frac{16Q^2}{M^2r^4}}{1-\frac{2R_s}{M^2r^3}+\frac{4Q^2}{M^2r^4}+\frac{4\Lambda}{3M^2}}\, .
 \end{eqnarray}
All three propagation speeds should be positive to guarantee the absence of laplacian instabilities. Again, at very large scales $r\gg r_{\rm eh}$ both propagation speeds become 1, as it corresponds to the de Sitter case. At intermediate scales well below the de Sitter horizon (so we can neglect the effects of the cosmological constant) but outside the event horizon $r>r_{\rm eh}$ we find the stability region shown in Fig. \ref{RN}, which has again been obtained by minimizing the corresponding propagation speeds with respect to $r$ for $r\geq r_{\rm eh}$. As it happened with the absence of ghosts conditions, the laplacian stability conditions barely depend on the black hole electric charge. Again, this originates from the second order correction introduced by the charge, as explained above.
 
In this section we have seen that the introduction of an electric charge for a black hole does not significantly modify the stability conditions found in the previous section for the Schwarzschild spacetime, so that all the bounds on the non-minimal Horndeski interaction also hold for the case of a charged black hole. One interesting consequence of our findings is that the static and spherically symmetric solutions for the full system of Einstein equations with the Horndeski interaction can be obtained perturbatively around the usual solution for Maxwell-Einstein equations because the stability of the theory requires the Horndeski term to remain small. Of course, one could also have additional solutions different from Reisner-Nordstr\"om as in \cite{MuellerHoissen:1988bp}, although a stability analysis would be required.

\section{Discussion \label{discussion}}
In this paper we have studied a general theory for a vector field with non-minimal couplings to gravity and  second order equations of motion. The resulting theory turns out to be surprisingly simple, consisting only of the usual gauge invariant Maxwell term and a quadratic coupling of the Faraday tensor to the dual of the Riemann tensor. This is the result that was obtained by Horndeski \cite{Horndeski:1976gi}. It is also interesting to note that the theory has only one free parameter.
We have studied this theory in several background spacetimes with particular attention on  its stability. In de Sitter, we have computed the energy density of the corresponding modes and we have shown that the energy of physical modes is positive for $1+2H^2/M^2>0$, which is guaranteed for $M^2>0$.

We have studied Bianchi I solutions generated by a cosmological constant in the presence of the vector field with the non-minimal coupling. We have shown that the isotropic de Sitter solution is an attractor of the phase map so that it corresponds to a stable solution. Interestingly, neither the position nor the stability properties of the critical points depend on the non-minimal coupling, i.e, on the value of $M^2$. Moreover, if the initial anisotropy of the universe is smaller than one, $R<1$, the system is always attracted towards the de Sitter solution and, even if the initial anisotropy is large but {\it negative}, this is the equilibrium solution. On the other hand, the region with $R>1$ has been shown to be unphysical because the vector field energy density is negative in that region.

We then have studied the behavior of the vector field with the non-minimal coupling for different classical solutions like  de Sitter, a Friedmann universe, Schwarzschild  and Reissner-Nordstr\"om-de Sitter space times. From observations we must require the stability of all solutions which we observe in nature. An unstable solution decays rapidly due to fluctuations and can therefore never be observed as final states. Thus, we have obtained the conditions for the absence of ghost-like and Laplacian instabilities for the mentioned physically relevant geometries. For all considered spacetimes we show that the Maxwell theory has a stable neighborhood. Quite generally, however, instabilities are present in the regime where the non-minimal interaction energy is of the same order or greater than the contribution from the minimal terms. 

Requiring stability around solar mass black holes yields a limits of about $|M|\gsim10^{-10}$eV$(M_{\odot}/M_*)$.  Since no black holes with $M_*\lsim M_{\odot}$ have been observed so far, this yields a limit  $|M|\gsim10^{-10}$eV. From the analysis of FLRW universes, stability requests $|M|^2 > 4H^2$ during a radiation dominated epoch (where $q=1$).  We know from Big Bang Nucleosynthesis (BBN) that the Universe has at least achieved a temperature
of $T_{BBN}\simeq 1$ MeV in the past which corresponds to a Hubble scale $H_{BBN} \simeq 1$sec$^{-1} \simeq 10^{-15}$eV. Hence $|M|>10^{-15}$eV.  This limit is less stringent than the one from black holes. If we assume that the temperature of the Universe achieved much higher values, say $T_*$ at the end of reheating after inflation, we obtain the constraint
\be
|M| > g_*^{1/2} \frac{T_*^2}{M_p} \,.
\ee
Here $g_*$ is the effective number of degrees of freedom after reheating. Depending on $T_*$, this limit can be quite interesting.

If the theory is viewed as a modification of electrodynamics so that the gauge field couples to charged fermions, which was Horndeski's perspective \cite{Horndeski:1976gi}, one must require $M\ll m_e$ where $m_e$ is the electron mass, to ensure that radiative corrections in QED \cite{Drummond} (which spoil the second order structure of the field equations) can be neglected.  Even in that case there is a huge parameter region $|M| \in (10^{-10}$ eV, $511$ keV) where the theory can be regarded as safe. The lower limit comes from the black hole bound derived in this paper and the upper limit is the electron mass.  If the theory is instead viewed as a hypothetical vector-tensor theory in which the vector does not represent the gauge boson of electrodynamics, the viable parameter range will be even larger because the upper limit should be replaced by the Planck mass to avoid radiative corrections from gravitons or by the mass of the lightest particle coupled to this gauge boson.

The fact that stability requires $|M^2| > 4H^2$ means that we can never be in the regime $H^2/M^2\gg 1$ where the energy density of a homogeneous vector field decays much slower than $1/a^4$ as we discussed in Section~\ref{sec:test}. So this type of non-minimal interactions does not represent a promising mechanism to maintain large scale magnetic fields at late times.

We have also calculated the propagation speed of the vector in the considered spacetimes.  We found that superluminality is a quite generic feature in this theory. In FLRW we showed that superluminality is avoided for $M^2>0$ if the null energy condition holds.  In Schwarzschild spacetime, however, superluminality is an unavoidable consequence of the non-minimal interaction because, regardless of the sign of $M^2$, there will always be some propagation direction and polarization for which the propagation speed is greater than $1$. Since the propagation speeds that we have computed do not depend on the Fourier mode and they all are massless (i.e., the dispersion relations can be written as $\omega^2=c_s^2k^2$ with $c_s^2$ independent of $k$), they also correspond to the group velocity.  Therefore, the propagation velocities we have calculated are the signal speeds measured by specific observers (comoving in FLRW and static in Schwarzschild). In Minkowski spacetime, it is easy to construct paradoxes concerning the causal structure of events forbidding faster-than-light transmission of information.  When curvature is involved, however, this is less obvious and it is sometimes argued that no general principle is violated by superluminal propagation \cite{Drummond}. This is not a feature that crucially depends on the non-trivial curvature, but it is a general property of hyperbolic systems when Lorentz violations are present \cite{Babichev:2007dw}.

In fact, superluminal propagation does not necessarily imply acausality.  Even in General Relativity one perfectly encounters valid solutions which allow closed timelike curves. However, as Hawking  showed in his Chronology Protection, the corresponding energy-momentum of these particles traversing the closed timelike curves becomes so large that its backreaction on the spacetime destroys the geodesic path \cite{Hawking:1991nk}. Similar Chronology Protection also occurs in Galileon type of theories \cite{Burrage:2011cr}.  Furthermore, quantum fields propagating in a curved background can lead to the appearance of superluminal propagation (as it happens in QED). However, a careful treatment of such quantum fields show that causality (and even unitarity) are actually not violated (see the extensive and detailed treatment in \cite{Hollowood:2011yh}) Therefore, unlike the conditions coming from ghosts and Laplacian instabilities, we have not used superluminality as a criterion for the viability of the theory.

\acknowledgments
This work is supported by the Swiss National Science Foundation. J.B.J. is supported by the Wallonia-Brussels Federation grant ARC No.~11/15-040 and also thanks support from the Spanish MICINN’s Consolider-Ingenio 2010 Programme under grant MultiDark CSD2009-00064 and project number FIS2011-23000. MT would like to thank CP3 at Universit\'e Catholique de Louvain where parts of this work was done for the hospitality.

\end{document}